\documentclass[a4paper,12pt]{article}
\usepackage{etex}
\usepackage{epsf}
\usepackage[dvips]{graphicx}
\usepackage{a4wide}
\usepackage{amsmath}
\usepackage{amscd}
\usepackage{amsfonts}
\usepackage{natbib}
\usepackage{multirow}

\usepackage[T1]{fontenc} 

\usepackage{microtype} 
\usepackage[hang, small,labelfont=bf,up,textfont=it,up,skip=2.5pt]{caption}
\usepackage{booktabs}
\usepackage{lettrine}

\usepackage{abstract} 

\usepackage{titlesec} 
\titleformat{\section}[block]{\large\scshape\centering}{\thesection.}{1em}{} 
\titleformat{\subsection}[block]{\large}{\thesubsection.}{1em}{} 

\usepackage{titling} 
\usepackage{amssymb}
\usepackage{caption,subfig}
\usepackage{longtable}
\usepackage{rotating}
\usepackage{wrapfig}
\usepackage{pdflscape}
\usepackage{color}
\usepackage{multirow}
\usepackage{rotating}
\usepackage{natbib}
\usepackage{enumerate}
\usepackage{enumitem}
\usepackage{dsfont}
\usepackage{setspace}
\usepackage{anysize}
\usepackage{booktabs}
\usepackage{soul}
\usepackage{float}
\usepackage{enumitem}
\usepackage{tikz}
\usepackage{pifont}
\usepackage[hidelinks]{hyperref}
\usepackage{moresize}
\usetikzlibrary{arrows}
\usepackage{multirow}
\usepackage{ntheorem}
\usepackage{geometry}
\usepackage{tabularx}
\hypersetup{colorlinks=true,urlcolor=blue,linkcolor=blue,citecolor=blue}
\usepackage[section]{placeins}
\usepackage{textcomp}
\newcommand{\sym}[1]{{#1}} 
\theoremstyle{break}
\newtheorem{theorem}{Proposition}

\newtheorem{proposition}[theorem]{Proposition}

\newtheorem{assumption}{Assumption}
\numberwithin{equation}{section}
 
\newenvironment{proof}[1][Proof]{\begin{trivlist}
\item[\hskip \labelsep {\bfseries #1}]}{\end{trivlist}}

\newcommand{\qed}{\nobreak \ifvmode \relax \else
      \ifdim\lastskip<1.5em \hskip-\lastskip
      \hskip1.5em plus0em minus1/2em \fi \nobreak
      \vrule height0.75em width1/2em depth0.25em\fi}
      
\marginsize{3cm}{3cm}{3cm}{3cm}
\long\def\symbolfootnote[#1]#2{\begingroup
\def\thefootnote{\fnsymbol{footnote}}\footnote[#1]{#2}\endgroup}

\DeclareGraphicsExtensions{.eps}

\date{}

\title{Peer Effects with Miss-specified Peer Groups}
\author{Christiern Rose (UQ), Lizi Yu (UQ)}

\renewcommand{\maketitlehookd}{%
\begin{abstract}
\noindent We consider identification of peer effects under peer group miss-specification. Two leading cases are missing data and peer group uncertainty. Missing data can take the form of some individuals being entirely absent from the data. The researcher need not have any information on missing individuals and need not even know that they are missing. We show that peer effects are nevertheless identifiable under mild restrictions on the probabilities of observing individuals, and propose a GMM estimator to estimate the peer effects. In practice this means that the researcher need only have access to an individual level sample with group identifiers. Group uncertainty arises when the relevant peer group for the outcome under study is unknown. We show that peer effects are nevertheless identifiable if the candidate groups are nested within one another and propose a non-linear least squares estimator. We conduct a Monte-Carlo experiment to demonstrate our identification results and the performance of the proposed estimators, and apply our method to study peer effects in the career decisions of junior lawyers.\\

\noindent\textbf{Key words:} Peer effects, missing data, measurement error, networks
\end{abstract}
}

\begin{document}


\maketitle

\onehalfspacing

\section{Introduction}

We address two challenges faced by researchers studying peer effects, each of which can lead to miss-specification of peer groups. The first challenge is that standard methods require that the researcher has access to \textit{group data}: a sample of groups which includes the outcome and characteristics of \textit{all} members (see \cite{bramoulle20} for a recent review). Such data are often proprietary or restricted-access \citep{breza20}, and widely available individual level survey data cannot be used. 
Without group data, empirical practice is either to drop individuals with missing peer data, leading to sample selection and loss of information, or to use only non-missing peers, leading to measurement error. Measurement error also arises if the researcher is unaware that peers are missing. We refer to this as the \textit{missing data} problem, and propose a solution which corrects for measurement error and makes full use of the available information.

The second challenge is that the researcher has to choose the relevant peer group, often from a set of candidate group structures. For example, studies based on the Dartmouth room-mate data, in which college freshman were randomly allocated to dorm rooms, have noted that it isn't clear whether peer effects operate at the room or the floor levels \citep{sacerdote01,glaeser03,angrist14}, and the relevant group may be different for different outcomes, such as academic attainment and fraternity membership \citep{sacerdote01}. 
Empirical practice is to conduct a robustness test by re-estimating the peer effects for each candidate group. This implicitly assumes that the relevant group is the same for all individuals (i.e., it is deterministic), but there is no reason for this to be the case. In the Dartmouth context, some dorms may simply be more sociable than others, and the researcher is unlikely to know which ones. Moreover, the researcher does not know which (if any) of the estimates are valid, which can be problematic when there are large differences in the estimates. 
We refer to this as the \textit{group uncertainty} problem, and propose a solution based on random peer group structure.



We first show that missing data and group uncertainty are examples of a wider class of miss-specification in which the members of the specified peer group are a \textit{subset} of the members of the (true) group. This is clear for missing data if the specified group comprises the observed group members. For group uncertainty, it is applicable in the typical empirical setting in which the candidate groups are nested and the specified group is the smallest group. For example, in the Dartmouth context rooms are nested within floors and groups can be specified based on rooms. 

Under subset-miss-specification, we show that peer effects can be identified using assumptions which allow the distribution of the group size to be inferred. 
With missing data, we show that widely used assumptions on the missingness mechanism allow the group size to be inferred by restricting the distribution of the group size conditional on the specified group size (i.e., on the number of observed individuals from that group). For group uncertainty, we suppose that there are two nested candidate groups (e.g., dorm rooms and floors), from which the relevant group is exogenously determined at random.\footnote{The extension to three or more candidate groups is straightforward.} This also restricts the distribution of the group size conditional on the sizes of the two candidate groups.

We apply our approach to settings with both small and medium-large group sizes. Our Monte-Carlo experiment is designed to mimic the Dartmouth room-mate data used in \cite{sacerdote01,glaeser03} and \cite{angrist14}, in which students were assigned to dorm rooms of size 2, 3 or 4. Our results demonstrate that the biases arising from ignoring missing data and group uncertainty can be large, but are corrected by the GMM and NLS estimators we propose. 

Our empirical application studies how peer ability impacts lawyers' decision to exit the local legal market using unique employer-employee matched data comprising all lawyers practicing in Shanghai, China. Peer groups are specified based on the number of junior lawyers practicing in the same firm, with a mean of 11.2 and a maximum of 482. We find that the likelihood of a lawyer's quit-to-exit increases in the proportion of high-ability peers, which is consistent with the invidious comparison model \citep{hoxby2005taking,antecol2016peer}. We apply our results on missing data to show that similar estimates and qualitative conclusions could have been reached had we only had access to an individual level sample of lawyers, rather than the group data we use. Finally, we apply our results on group uncertainty to study whether peer effects operate at the firm or firm-cohort levels, and find evidence of considerable heterogeneity in the relevant peer group across firms.
 
\subsection{Related literature}\label{lit}

The literature on sampled networks, mismeasured networks and missing data has recently attracted much interest. \cite{lewbel191} and \cite{hardy19} consider the implications of measurement error in the network through which peer effects operate. \cite{lewbel191} also show that peer effects can be identified and consistently estimated when there is no information on the network beyond group identifiers. \cite{boucher20} study peer effects when the network is unknown but that its distribution is known or can be consistently estimated. 

\cite{chand11} studies the implications of using sampled network data in applied work, provide analytical corrections in some examples of interest and develop a more general graphical reconstruction approach. In the context of peer effects, the analytical correction can be applied if the researcher has data on the network, outcomes and exogenous characteristics for a subsample of individuals and knows the identities, outcomes and exogenous characteristics of all of their peers. This means that data are only missing for peers-of-peers. To apply graphical reconstruction, for every individual, the researcher must observe the outcomes, exogenous characteristics and variables which can be used to predict the propensity of individuals to form links in a network formation model. 

\cite{breza20} use a parametric network formation model to show that network structure can be identified and used to construct network statistics (e.g., centrality measures) if the researcher does not collect network data for all individuals, but instead collects relational data for a sub-sample of individuals and basic characteristics of all individuals.\footnote{Relational data are collected by asking questions of the form `Consider all the individuals in your group with whom you do activity X. How many of these have trait Y?' \citep{breza20}} \cite{sojourner13}, \cite{wang13} and \cite{liu17} consider the case in which the network and all individuals are observed without error but there are missing data in the outcomes and/or exogenous characteristics. 

All of the above papers suppose that the researcher observes at least some data on every individual. In contrast, we allow for some individuals to be entirely missing and do not require knowledge on whether there even exist such individuals. These gains come from exploiting the structure of the group interactions model we consider, which, though widely used in practice (e.g., \cite{moffitt01,sacerdote01,glaeser03,angrist14,boucher14,cornel17}) and well understood theoretically (e.g., \cite{lee07,bramoulle09,davezies09,bramoulle20}), is less general than some.\footnote{We discuss this issue further in Section \ref{disc}.} 

Our work is also related to the literature on multiple peer effects. \cite{goldsmith13}, \cite{arduini20} and \cite{reza21} study a setting in which individuals are exposed to multiple peer effects, each operating through a different network. 
In contrast, we suppose that individuals are exposed to peer effects operating through one network, but there is uncertainty as to which is the relevant network. Our model is thus not a model of multiple peer effects, but one of peer effects with uncertain peer groups. The approaches are not observationally equivalent. With multiple peer effects, the endogenous peer effects propogate through a network formed by taking the union of the links over multiple networks. In contrast, with uncertain peer groups, the endogenous effects propogate through only one network, the identity of which is unknown.

Our empirical application contributes to the literature on peer effects in competitive environments, particularly with respect to peers' ability impact on lower- and medium-achieving individuals. One strand of this literature focuses on sports competitions and tournaments and mostly find a negative impact. \cite{brown2011quitters}, \cite{smith2013peers} and \cite{emerson2018peer} find a negative peer ability effect, in contrast to \cite{guryan2009peer} who find no effect. \cite{yamane2015peer} find positive effects when subjects are ahead and negative effects when they are behind. Another strand of this literature focuses on educational outcomes and reports mixed findings. Some studies find significant peer effects (for example, \cite{hoxby00} and \cite{sacerdote01} find a positive effect and \cite{antecol2016peer} find negative effect), and others find small to no effects (for example, \cite{foster2006s} and \cite{angrist2004does}). \cite{brady2017bad} find negative effects at broader level but positive effects at smaller level, suggesting that peer effects operate differently for different groups. The overall perception is that effects are not identical across different ability groups. Low-achieving groups are adversely affected by peer ability and may benefit from tracking compared to ability-mixing groups \citep{duflo2011pee,carrell2013natural,booij2017ability}. Our work builds on these papers by examining career decisions in the highly competitive and incentivized labor market for junior lawyers. Our results are more in line with the literature finding peer ability hurts subjects' persistence in education and in their chosen career \cite{thiemann2022persistent, howell2021learning}.

\section{Model}\label{mode}

We consider a group interactions model in which the data comprise an i.i.d. sample of $G_0$ groups. The asymptotic is $G_0\to\infty$, which is analagous to the large-$N$ fixed-$T$ asymptotic used for microeconometric panel data \citep{bramoulle09}. Individual $i\in\{1,2,...,N_0\}$ is in exactly one group denoted $g_0(i)\in\{1,2,...,G_0\}$. The size of $i$'s group is $n_{g_0(i)}\geq 2$. When it is not required, we omit the dependence on $i$ and simply refer to $g_0$ and $n_{g_0}$. The continuous outcome $y_{i}$ of individual $i$ is determined by
\begin{align}
y_{i}=\alpha_{g_0}+\left(\frac{1}{n_{g_0}-1}\sum_{\substack{j:g_0(j)=g_0(i)\\j\neq i}}y_{j} \right)\beta+\left(\frac{1}{n_{g_0}-1}\sum_{\substack{j:g_0(j)=g_0(i)\\j\neq i}}x_{j} \right)\delta+ x_{i}\gamma+\epsilon_{i},\label{struc}
\end{align}
where $\alpha_{g_0}$ is group unobserved heterogeneity, $x_{i}$ is an exogenous characteristic and $\epsilon_{i}$ is an error term satisfying 
\begin{align}
\mathbb{E}[\epsilon_{i}|\alpha_{g_0},n_{g_0},\mathbf{x}_{g_0}]=0\quad \forall i\in\{1,2,...,N_0\},\label{mom}
\end{align} 
where $\mathbf{x}_{g_0}=(x_{j})_{j:g_0(j)=g_0(i)}$ is $n_{g_0}\times 1$.\footnote{Since we only consider the within-transformation of the fixed effects model below, \eqref{mom} can be relaxed to $\mathbb{E}[\epsilon_{i}|n_{g_0},\mathbf{x}_{g_0}]=0$. We maintain the form of \eqref{mom} to maintain comparability with the literature.} The moment condition supposes that group size and the characteristics of the group members are exogenous with respect to $\epsilon_{i}$. However, both can be arbitrarily dependent on $\alpha_{g_0}$, which allows for selection of individuals into groups. The model includes correlated effects (captured by $\alpha_{g_0}$), endogenous peer effects ($\beta$) and contextual peer effects ($\delta$).  

For simplicity of exposition, following \cite{bramoulle09} we present results for the case where there is a single characteristic $x_{i}$. All results continue to apply in the more general case in which $x_{i}$, $\gamma$ and $\delta$ are vectors. As explained in Section \ref{lit}, this model is well understood from a theoretical perspective and is pervasive in applied work. Moreover, it can be microfounded based on a game in which utility depends on one's own action and the average action of the others in the group (see \cite{blume15,bramoulle20}).

To ensure that the reduced form of \eqref{struc} exists, as is standard in the literature, we suppose that $|\beta|<1$ (i.e., that the endogenous peer effect is not explosive), yielding
\begin{align}
y_{i}&=\frac{\alpha_{g_0}}{1-\beta}+ \left(\sum_{\substack{j:g_0(j)=g_0(i)\\j\neq i}}x_{j} \right)\pi_{1}(n_{g_0})+ x_{i}\pi_{2}(n_{g_0})+u_i,\label{red}\\
\pi_{1}(n)&\triangleq\left(\frac{(\delta+\beta\gamma)(n-1)^{-1}}{1-\beta(\beta+n-2)(n-1)^{-1}}  \right),\quad \pi_{2}(n)\triangleq\left(\frac{\gamma+\beta(\delta-\gamma(n-2))(n-1)^{-1}}{1-\beta(\beta+n-2)(n-1)^{-1}}\right),\nonumber
\end{align}
where $u_i$ is the reduced form error. Applying the within-group transformation, $$\widetilde{y}_{i}\triangleq y_{i}-\frac{1}{n_{g_0}}\sum_{j:g_0(j)=g_0(i)}y_{j},$$ to the reduced form yields,
\begin{align}
\widetilde{y}_{i}= \widetilde{x}_{i}\pi(n_{g_0})+\widetilde{u}_i\label{red2},\quad\pi(n)\triangleq\left(\frac{\gamma-\delta(n-1)^{-1}}{1+\beta(n-1)^{-1}}\right).
\end{align}
Though other transformations can be used, we focus on the within-group transformation since it delivers the least information loss (see Section 3.3 of \cite{bramoulle09}).

\section{(Possibly) Miss-specified Groups}\label{msg}

The researcher specifies the model in Section \ref{mode} with group $g(i)\in\{1,...,G\}$ for individual $i$. The size of group $g(i)$ is $n_{g(i)}$. As with $g_0(i)$, we omit the dependence of $g(i)$ on $i$ unless it is required. The specified groups may cover only a subset $\mathcal{S}\subseteq\{1,...,N_0\}$ of individuals of cardinality $N\leq N_0$. This could arise, for example, if the available data comprise a random sample of the $N_0$ individuals. However, we maintain that every individual in $\mathcal{S}$ is in exactly one specified group. We also make use of the within-\textit{specified}-group transformation, $$\overline{y}_{i}\triangleq y_{i}-\frac{1}{n_{g}}\sum_{j:g(j)=g(i)}y_{j},$$
which would be applied to eliminate specified group fixed effects. To build intuition on the implications of miss-specification for identification, we now consider two examples and derive their implications for the reduced form, which will later be used for our identification analysis. \vspace{0.25cm}

\noindent \textbf{Example 1 - Superset miss-specification.} Suppose that two groups, each of size $n_0/2$, are combined to form a specified group of size $n_0$.  Solving for the reduced form and applying the within-specified-group transformation yields, for individual $i$ in group 1
\begin{align}
\overline{y}_i
&=\overline{x}_i\pi(n_0)+(a+b+c+\overline{u}_i)\label{redex}\\
&=\overline{x}_i\pi(n_0)+\overline{v}_i\label{redex1},
\end{align}
where,
\begin{align*}
&a=\frac{\alpha_{1}-\alpha_{2}}{2(1-\beta)},\quad b=\left(\frac{\gamma+\delta}{1-\beta}\right)\frac{1}{n_0}\left(\sum_{j:g_0(j)=1}x_j-\sum_{j:g_0(j)=2}x_j\right),\\&c=-\frac{\overline{x}_in_0(\gamma\beta+\delta)}{2(n_0-1+\beta)(n_0/2-1+\beta)}.
\end{align*}
There are three sources of miss-specification, each corresponding to a term in equation \eqref{redex}. First, the researcher miss-specifies the correlated effect, yielding the term $a$. Second, by specifying only one group the researcher erroneously includes the exogenous characteristics and outcomes of those in group 2 in the structural equation determining the outcomes for group 1. This yields the term $b$. Finally, the researcher miss-specifies the group size to be $n_0$ instead of $n_0/2$. This leads the reduced form parameter to be incorrectly specified as $\pi(n_0)$ rather than $\pi(n_0/2)$, yielding the term $c$. Due to these terms, the error $\overline{v}_i$ depends on $\overline{x}_i$ and $n_0$.
\vspace{0.25cm}

\noindent \textbf{Example 2 - Subset miss-specification.} Now switch the roles of the groups and the specified groups in Example 1, such that the researcher divides one group into two specified groups, each of size $n_0/2$. Then we obtain, for individual $i$ in group 1,
\begin{align}
\overline{y}_i
&=\overline{x}_i\pi(n_{0}/2)+(c+\overline{u}_i)\label{redex2}\\
&=\overline{x}_i\pi(n_{0}/2)+\overline{v}_i\label{redex3},
\end{align}
where this time,
\begin{align*}
c=\frac{\overline{x}_in_{0}(\gamma\beta+\delta)}{2(n_{n_0}/2-1+\beta)(n_{0}-1+\beta)}.
\end{align*}
Terms such as $a$ and $b$ in Example 1 do not arise. This is because all members of specified group 1 have the same correlated effect as those in specified group 2, and, the sum of the outcomes and exogenous characteristics over all members of specified group 2 (respectively,  specified group 1) is common to all members of specified group 1 (respectively,  specified group 2). The within-specified-group transformation thus eliminates both sources of miss-specification. Miss-specification only manifests through the group size (i.e., through $c$).
\vspace{0.25cm}

In Example 1 the specified group members are a \textit{superset} of the group members, and there are three sources of miss-specification. In Example 2 they are a \textit{subset}, and there is one source of miss-specification.\footnote{Though we do not present such an example, the case in which we have neither a subset nor a superset is similar to the superset case.}  Examples 1-2 suggest that subset miss-specification is more readily addressed because one only needs to be concerned with specification of the group size (i.e., the term $c$).  This is important, because there is no clear solution to terms such as $a$ and $b$ without making strong restrictions on the distributions of $\mathbf{x}_{g_0}$ and $\alpha_{g_0}$. As we show below, the intuition that subset miss-specification is more easily addressed is not specific to Examples 1-2. We also argue below that subset miss-specification is more relevant in practice. We thus proceed under the following assumption,
\begin{assumption}[Subset miss-specification]\label{gnest}
For all $(i,j)\in\mathcal{S}^2$, $g(i)=g(j)\Rightarrow g_0(i)=g_0(j)$,
\end{assumption}
and the remainder of the paper focuses on miss-specification arising due to incorrect specification of the group size. 

Assumption \ref{gnest} covers the case of missing data by allowing the researcher to observe $N<N_0$ individuals with outcomes, characteristics and group identifiers. For example, in the context of education, the researcher might access a sample of students' test-scores, characteristics and classroom/school identifiers (e.g., \cite{davezies09}) or have access to a large educational survey such as the student level PISA survey, which contains school and grade identifiers but includes only a subset of students.\footnote{Other examples include risky behaviours and neighbourhood effects. For the former, survey data on smoking, drinking and illicit drug use among students can be incomplete \citep{lundborg06}. For the latter, the neighbourhood average outcome may be measured using survey data which includes geographic identifiers (e.g., census tract). For example, \cite{bertrand00} use the 5\% public use microsample of the 1990 Census to study welfare take-up. \cite{glaeser03} use the same data for wages.} 

Assumption \ref{gnest} also covers the case in which there is group uncertainty but the candidate groups are nested. It is satisfied provided that the researcher uses the smallest group structure to specify $g$. For example, \cite{glaeser03} and \cite{angrist14} consider peer groups based on dorm rooms and floors,\footnote{Similarly, peer effects in education may operate at the classroom, grade or school levels. Neighbourhood effects may operate at at the two, three or four digit postcode level.} so Assumption \ref{gnest} holds if the specified groups are rooms. In our empirical application, we postulate that peer effects among lawyers may operate either at the firm or firm-cohort levels, in which case Assumption \ref{gnest} holds if firm-cohorts are specified.

\section{Identification}\label{iden}

We now study identification under Assumption \ref{gnest}. Our identification analysis follows \cite{bramoulle09}, hence we say that the structural parameters are (point) identified if and only if they can be uniquely recovered from the reduced form parameters presented below (i.e., there is an injective relationship). Our results are thus asymptotic in nature (see \cite{manski95}), and characterize whether endogenous, contextual and correlated effects can be distentangled if there is no limit to the number of groups $G_0$.

Under Assumption \ref{gnest}, the reduced form for individual $i\in\mathcal{S}$ is
\begin{align}
\overline{y}_{i}&=\overline{x}_{i}\pi(n_{g_0})+\overline{u}_i\label{red3}
\end{align}
and taking conditional expectations yields
\begin{align}
\mathbb{E}[\overline{y}_{i}|n_{g},\mathbf{x}_{g}]&= \overline{x}_{i}\varphi(n_g,\mathbf{x}_g)+\mathbb{E}[\overline{u}_i|n_{g},\mathbf{x}_{g}],\label{redE}
\end{align}
where $\varphi(n,\mathbf{x})\triangleq \mathbb{E}[\pi(n_{g_0})|n_{g}=n,\mathbf{x}_{g}=\mathbf{x}]$. To ease the notational burden we do not make explicit that the expectations are also conditional on $i\in\mathcal{S}$. Below, we consider only cases in which this omission is innocuous and omit the condition $i\in\mathcal{S}$ for the remainder of the paper. 

Equation \eqref{redE} shows that identification prospects depend on whether there is within-specified-group variation in the exogenous characteristic, on whether $\mathbb{E}[\overline{u}_i|n_{g},\mathbf{x}_{g}]=0$, and if both of these conditions are satisfied, on whether the structural parameters can be uniquely recovered from the reduced form parameters $\varphi(n,\mathbf{x})$, which are identified for all $(\mathbf{x},n\geq 2)$ in the support of $(\mathbf{x}_{g},n_g)$.\footnote{We require $n\geq 2$ because $\varphi(1,\mathbf{x})$ is not identifiable since $n_g=1$ implies $\overline{y}_i=\overline{x}_i=0$.}

In general, even under Assumption \ref{gnest}, the moment condition in \eqref{mom} does not imply $\mathbb{E}[\overline{u}_i|n_{g},\mathbf{x}_{g}]=0$. Suppose however that we can make a restriction such that it is satisfied, and also that there is within-specified-group variation in the exogenous covariate. Then the structural parameters are point identified if they can be uniquely recovered from the reduced form parameters $\varphi(n,\mathbf{x})$ and the equations
\begin{align}
\varphi(n,\mathbf{x})=\sum_{m=2}^\infty\mathbb{P}[n_{g_0}=m|n_{g}=n,\mathbf{x}_{g}=\mathbf{x}]\pi(m)\label{ide}
\end{align}
for all $(\mathbf{x},n\geq 2)$ in the support of $(\mathbf{x}_{g},n_g)$. The distribution $n_{g_0}|n_{g},\mathbf{x}_{g}$ is not observed, hence the structural parameters are not point identified unless it can be restricted.\footnote{Partial identification is possible because the probabilities $\mathbb{P}[n_{g_0}=m|n_{g},\mathbf{x}_g]$ for $m=2,3....$ are non-negative and sum to 1. We do not pursue partial identification, since, as shown below, typical empirical settings faced by researchers can lead to point identification.} 

We now consider three assumptions, each of which guarantees $\mathbb{E}[\overline{u}_i|n_{g},\mathbf{x}_{g}]=0$ and restricts the distribution of $n_{g_0}|n_{g},\mathbf{x}_{g}$. Each assumption includes the standard model of peer effects with known peer groups as a special case, respectively when $n_g=n_{g_0}$ (Assumption \ref{ass1}), when $\rho=1$ (Assumption \ref{ass2}) and when $\psi\in\{0,1\}$ (Assumption \ref{ass3}). 

To rule out pathological cases and ease the notational burden, we present our our results for the case in which there is within-specified-group variation in the exogenous characteristic, and that the support of the distribution of $n_{g_0}|\mathbf{x}_{g_0}$ does not depend on $\mathbf{x}_{g_0}$. The results in Propositions \ref{the1} and \ref{the2} continue to apply if the latter is relaxed, provided that there exists an element in the support of the exogenous characteristic such that the support condition holds for the conditional distribution of $n_{g_0}$.

\subsection{Missing data}

We first consider identification when there are missing data. Our analysis makes uses of the indicator $s_i\in\{0,1\}$ for the condition $i\in\mathcal{S}$.

\begin{assumption}[Known group size]\label{ass1}
For each individual $i\in\{1,2,...,N_0\}$, the researcher observes $(y_{i},x_{i},g_0(i),n_{g_0(i)})$ when $s_i=1$ and $\mathbb{E}[\epsilon_i|\alpha_{g_0},n_{g_0},\mathbf{x}_{g_0},\mathbf{s}_{g_0}]=0$. The researcher groups individuals by $g_0$.
\end{assumption}
Assumption \ref{ass1} covers the simplest case in which the group size is known but individuals are sampled from the group. In practice, it means that the researcher has access to a sample of individuals with outcomes, characteristics, group membership indicators and group size. The distribution of $s_i$ (i.e., the inclusion probability) can be heterogeneous across individuals, but ought not to depend on the structural error (i.e., there should be no sample selection on unobservables). Identification under Assumption \ref{ass1} was first considered by \cite{davezies09}. We include it for comparison with our approach, which allows the group size to be unknown.
\begin{assumption}[Unknown group size]\label{ass2}
For each individual $i\in\{1,2,...,N_0\}$, the researcher observes $(y_{i},x_{i},g_0(i))$ when $s_i=1$, $\mathbb{E}[s_i|\mathbf{x}_{g_0},n_{g_0}]=\rho\in(0,1]$, $\mathbb{COV}[s_i,s_j|\mathbf{x}_{g_0},n_{g_0}]=0$ for $j\in g_0(i), j\neq i$, and $\mathbb{E}[\epsilon_i|\alpha_{g_0},n_{g_0},\mathbf{x}_{g_0},\mathbf{s}_{g_0}]=0$. The researcher groups individuals by $g_0$.
\end{assumption}
Assumption \ref{ass2} relaxes the requirement that the researcher knows the group size. In practice, it allows for the case in which the researcher observes an individual level sample of outcomes, exogenous characteristics and group membership indicators. In such a sample, the researcher knows the number of individuals sampled from each group ($n_{g}$) but does not know the group size ($n_{g_0}$), nor whether there are any missing individuals in each group. Relative to Assumption \ref{ass1}, we additionally assume that the sample of observed individuals is unbiased and uncorrelated. These are mild assumptions likely to be satisfied in well designed individual level samples under common sampling schemes \citep{sampling}.

Assumptions \ref{ass1} and \ref{ass2} both maintain that the sampling of individuals does not induce sample selection, whilst Assumption \ref{ass2} additionally maintains that the sampling does not depend on exogenous covariates nor on the group size. Viewed from the missing data perspective, Assumption \ref{ass2} is essentially a type of Missing Completely at Random assumption, which is widely used (often implicitly) in empirical work. Together, $\mathbb{E}[s_i|\mathbf{x}_{g_0},n_{g_0}]=\rho\in(0,1]$ and $\mathbb{COV}[s_i,s_j|\mathbf{x}_{g_0},n_{g_0}]=0$ imply

\begin{align}n_{g}|n_{g_0},\mathbf{x}_{g}\sim \text{Binomial}(n_{g_0},\rho),\label{bin}\end{align}
which leads to identification of the distribution $n_{g_0}|n_g,\mathbf{x}_g$ from the observed distribution $n_g|n_g\geq 1,\mathbf{x}_g$.\footnote{Conditioning on $n_{g}\geq 1$ is important, because for some groups the researcher may not observe any individuals. Hence it is the distribution $n_{g}|n_{g}\geq 1,\mathbf{x}_g$ which is observed rather than $n_{g}|\mathbf{x}_g$.}

We now briefly discuss possible generalizations of Assumption \ref{ass2}, and the extent to which they facilitate identification. First, one might consider $\rho=\rho_{g_0}$ to allow for group heterogeneity. Such an assumption does not provide information to identify $n_{g_0}|n_g,\mathbf{x}_g$ because $\mathbb{P}[n_{g_0}=m|\mathbf{x}_{g}]$ and $\rho_{g_0}$ only appear multiplicatively in the expression for $\mathbb{P}[n_{g}=n|n_{g}\geq 1, \mathbf{x}_{g}]$ for all $m$ and $n$. For the same reason, using $\rho=\rho(n_{g_0})$ does not provide identifying information. 

Alternatively one might consider $\rho=\rho(z_i)$ for some exogenous observable $z_i$ (e.g., $z_i=x_i$). Viewed from the missing data perspective, this corresponds to a Missing At Random assumption. In this case $n_{g}|n_{g_0},\mathbf{x}_{g_0},\mathbf{z}_{g_0}$ follows a Poisson Binomial distribution, which could, in principle, be used for identification. However, if $\mathbf{z}_{g_0}$ is observed then $n_{g_0}$ is known and identification is attained instead under the weaker Assumption \ref{ass1}. In contrast, the distribution $n_{g}|n_{g_0},\mathbf{x}_{g},\mathbf{z}_{g}$ does not take a known form, hence is of limited use for the identification in the typical setting in which we have no information on the missing individuals (i.e., there are some individuals for whom $z_i$ is unobserved). 

We do not pursue the above generalisations formally because the focus of our analysis is not on sample selection at the individual level, but instead on replacing a group level sample with an individual level sample. Hence we consider Assumption \ref{ass2} as the benchmark case in which the researcher has access to a well designed sample of individuals rather than a well designed sample of groups. We now present our identification result.
\begin{proposition}\label{the1}
$(\gamma,\beta,\delta)$ are point identified if $\gamma\beta+\delta\neq 0$ and any one of the following holds

\begin{enumerate}

\item Assumption \ref{ass1} holds and the support of the distribution of $n_{g_0}$ has at least three elements \citep{davezies09}.

\item Assumption \ref{ass2} holds and the support of the distribution of $n_{g_0}$ is bounded and has at least three elements.



\end{enumerate}

\end{proposition}

\noindent Proposition \ref{the1} shows that the structural parameters are identifiable if $\gamma\beta+\delta\neq 0$. This is a well known necessary identification condition in models with correctly specified groups and both endogenous and contextual effects (see for example \cite{bramoulle09} or \cite{rose17}), which is generically satisfied over the parameter space. If it is violated then the endogenous and contextual effects exactly offset one another such that the reduced form effect is $\pi(n)=\gamma$, which does not vary with $n$, and hence no amount of variation in group sizes can identify $\beta$ and $\delta$. Both results in Proposition \ref{the1} also require that the support of the distribution of $n_{g_0}$ has at least three elements. This is also necessary condition for identification when the groups are correctly specified and there are endogenous, contextual and correlated effects \citep{lee07,davezies09,bramoulle09}, hence it is also necessary if the groups are (possibly) miss-specified. 

Result 1 was first established by \cite{davezies09} for missing data with known group sizes whereas result 2 applies when the group sizes are unknown. Point identification of $\rho$ comes in part from information on the lower bound of the support of $n_{g_0}$. The result uses that $n_{g_0}\geq 2$, which is maintained throughout the theoretical and empirical literature (e.g., \cite{lee07,bramoulle09,davezies09,bramoulle20,moffitt01,sacerdote01,glaeser03,angrist14,boucher14,cornel17}). Indeed, groups of size one cannot provide information on peer effects and are not consistent with our model. However, if individuals are missing, then we may \textit{observe} groups with only one member even though all groups have at least two members (i.e., $\mathbb{P}[n_g=1|\mathbf{x}_g,n_{g}\geq 1]>0$ if $\rho<1$). Hence the distribution of $n_g$ provides information on $\rho$, which combined with \eqref{bin}, leads to identification of $\rho$ and the distribution $n_{g_0}|n_g,\mathbf{x}_g$. Equation \eqref{ide} can then be used to identify the peer effects. 

In some applications a stronger support restriction may be used if it is known that $n_{g_0}\geq \underline{n}$ for some known $\underline{n}>2$. For example, this is likely to be satisfied in studies of peer effects in education, where classrooms might reasonably be assumed to contain at least a handful of students. Though not required for identification, this can improve the precision of the GMM estimator we propose below. In other applications $n_{g_0}=1$ may be feasible, in which case one can never rule out $\rho=1$ and $n_{g_0}=n_g$ so that peer effects are not identifiable. For this reason, a restriction on the lower bound of the support of $n_{g_0}$ is necessary for identification of the peer effects.

Result 2 also uses the mild assumption that the distribution of $n_{g_0}$ is bounded, which is needed to identify $\rho$ and the distribution of $n_{g_0}|\mathbf{x}_g$ from the distribution of $n_g|n_g\geq 1,\mathbf{x}_g$ as a solution of a non-linear system comprising a finite number of equations. The upper bound need not be known since it is identifiable from the distribution of $n_g|n_g\geq 1$. This assumption is also used, for example, by \cite{lee07}, \cite{davezies09}, \cite{lewbel19} and \cite{boucher20}. 

\subsection{Group uncertainty}

We now consider subset miss-specification resulting from group uncertainty.
\begin{assumption}[Group uncertainty]\label{ass3}
For each individual $i\in\{1,2,...,N_0\}$ the researcher observes $(y_{i},x_{i},g_1(i),g_2(i),n_{g_1(i)},n_{g_2(i)})$, $2\leq n_{g_1}\leq n_{g_2}$ and $g_1(i)=g_1(j)\Rightarrow g_2(i)=g_2(j)$ for all $(i,j)\in\{1,2,...,N_0\}^2$. The researcher groups individuals by $g_1$, $\mathbb{E}[\epsilon_{i}|n_{g_1},n_{g_2},\mathbf{x}_{g_2}]=0$ and $\mathbb{P}[n_{g_0}=n_{g_1}|n_{g_1},n_{g_2},\mathbf{x}_{g_2}]=\psi \in[0,1]$ and $\mathbb{P}[n_{g_0}=n_{g_2}|n_{g_1},n_{g_2},\mathbf{x}_{g_2}]=1-\psi$.
\end{assumption}
Assumption \ref{ass3} covers the case in which there is group uncertainty but the candidate groups are naturally nested. Though an extension to three or more is straightforward, for simplicity of exposition we do not pursue it formally. 
We also require a minor strengthening of the moment condition, replacing \eqref{mom} with $\mathbb{E}[\epsilon_{i}|n_{g_1},n_{g_2},\mathbf{x}_{g_2}]=0$, and hence $n_{g}$ with $n_{g_1},n_{g_2}$ in equations \eqref{redE} and \eqref{ide}. The stronger moment condition requires that both potential group sizes and the exogenous characteristic of all individuals in the two potential groups be strictly exogenous with respect to the structural error. 

\begin{figure}[t!]
	\centering
	\caption{A connected component of $\mathcal{G}_{n_{g_1},n_{g_2}}$ with five vertices}    \label{fig:G}
    \includegraphics[height=5cm]{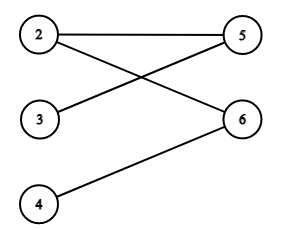}
 \begin{flushleft}
        \footnotesize{\textbf{Notes:} The vertices in the left hand column corresponds to $n_{g_1}$, which has support $\{2,3,4\}$. The vertices in the right hand column correspond to $n_{g_2}$, which has support $\{5,6\}$. An edge between $n_1$ and $n_2$ implies that $(n_1,n_2)$ lies in the support of $(n_{g_1},n_{g_2})$. This example satisfies the identification condition in Proposition \ref{the2} becase it is connected and has three vertices in the left hand column.}
\end{flushleft}
\end{figure}

Under Assumption \ref{ass3}, the reduced form is
\begin{align}
\mathbb{E}[\overline{y}_i|n_{g_1},n_{g_2},\mathbf{x}_{g_2}]=\overline{x}_i\varphi'(n_1,n_2,\mathbf{x})
\end{align}
where $\varphi'(n_1,n_2,\mathbf{x})\triangleq \mathbb{E}[\pi(n_{g_0})|n_{g_1}=n_1,n_{g_2}=n_2,\mathbf{x}_{g_2}=\mathbf{x}]=\psi\pi(n_{1})+(1-\psi)\pi(n_{2})$, hence identification hinges on the properties of the joint distribution of $n_{g_1}$ and $n_{g_2}$, specifically, through the bi-partite graph of the support of $(n_{g_1},n_{g_2})$, which we denote by $\mathcal{G}_{n_{g_1},n_{g_2}}$. Figure \ref{fig:G} depicts an example. 

Our approach can be generalized to allow $\psi=\psi(\mathbf{z}_{g_2})$ (e.g., $\mathbf{z}_{g_2}=\mathbf{x}_{g_2}$), provided that the conditional expectations and probabilities in Assumption \ref{ass3} continue to hold conditional also on $\mathbf{z}_{g_2}$. We discuss the corresponding modification of Proposition \ref{the2} at the end of this section. We now present our identification result.

\begin{proposition}\label{the2}
$(\gamma,\beta,\delta)$ are point identified if $\gamma\beta+\delta\neq 0$, Assumption \ref{ass3} holds, the support of the distribution of $n_{g_1}$ has at least three elements, the support of the distribution of $n_{g_2}$ has at least three elements and \textbf{either} there is a connected component of $\mathcal{G}_{n_{g_1},n_{g_2}}$ with at least three vertices corresponding to the support of $n_{g_1}$ \textbf{or} there is a connected component of $\mathcal{G}_{n_{g_1},n_{g_2}}$ with at least three vertices corresponding to the support of $n_{g_2}$.
\end{proposition}

\noindent The conditions $\gamma\beta+\delta\neq 0$ and that the supports of the distributions of $n_{g_1}$ and $n_{g_2}$ each have least three elements are necessary even when there is no miss-specification (i.e, when it is known that either $\psi=0$ or $\psi=1$) \citep{lee07,davezies09,bramoulle09}. In addition to these, we require a mild condition on the joint support of the candidate group sizes $(n_{g_1},n_{g_2})$, expressed through the bi-partite graph $\mathcal{G}_{n_{g_1},n_{g_2}}$. When this condition is satisfied, using standard arguments for bi-partite fixed effects models  (e.g., \cite{abowd99}), we are able to identify $\psi\pi(n_1)+(1-\psi)\pi(n_2)$ for all $(n_1,n_2)$ in a subset of the support of $(n_{g_1},n_{g_2})$. This subset is given by all pairs which appear in the same connected component of $\mathcal{G}_{n_{g_1},n_{g_2}}$. Importantly, there is no requirement that the supports of $n_{g_1}$ and $n_{g_2}$ overlap. Figure \ref{fig:G} provides such an example.

The intuition for identification of $\psi\pi(n_1)+(1-\psi)\pi(n_2)$ comes from a regression of $\overline{y}_i$ on the interaction of $\overline{x}_i$ with dummies for each value of $n_{g_1}$ and the interaction of $\overline{x}_i$ with dummies for each value of $n_{g_2}$. Together, both sets of dummies sum to two for every individual in the sample (because each individual is in exactly two candidate groups, a small one and a larger one), hence one dummy must be omitted. For this reason, we identify the pairwise sums of $\psi\pi(n_1)$ and $(1-\psi)\pi(n_2)$ for all pairs in the same connected component. If we observe a connected component containing at least three vertices corresponding to $n_{g_1}$ we can contrast $\psi\pi(n_{g_1})$ over three different values of $n_{g_1}$ holding $n_{g_2}$ constant. This yields three equations in the four unknowns $\psi,\gamma,\delta,\beta$. We obtain an additional equation by repeating the exercise for a different value of $n_{g_2}$, which leads to identification.

Under the generalization of Assumption \ref{ass3} to $\psi=\psi(\mathbf{z}_{g_2})$, Proposition 2 holds provided that the support conditions on $n_{g_1}$ and $n_{g_2}$ hold conditionally on $\mathbf{z}_{g_2}=\mathbf{z}$ for some element $\mathbf{z}$ of the support of $\mathbf{z}_{g_2}$. Similarly, one replaces $\mathcal{G}_{n_{g_1},n_{g_2}}$ by $\mathcal{G}_{n_{g_1},n_{g_2}|\mathbf{z}_{g_2}=\mathbf{z}}$. This rules out pathological cases such as $z_i=n_{g_2(i)}$.

\section{Estimation}\label{esti}

For missing data with known group sizes, we estimate $\gamma,\delta,\beta$ by non-linear least squares based on \begin{align}\mathbb{E}[\overline{y}_{i}|n_{g_0},\mathbf{x}_{g}]=\overline{x}_{i}\pi(n_{g_0}).\end{align}
We focus here on the non-linear least squares estimator since it is more easily adapted to the case of unknown group sizes. We present an alternative instrumental variables estimator in the appendix. If the group sizes are unknown, we use
\begin{align}\mathbb{E}[\overline{y}_{i}|n_{g},\mathbf{x}_{g}]=\overline{x}_{i}\sum_{m=2}^\infty\mathbf{p}_m(n_g;\rho,\mathbf{q}) \pi(m),\end{align} 
where $\rho\in(0,1]$, $0\leq\mathbf{q}_m\leq 1$ for $m=1,2,...$, $\mathbf{q}_1=0$, $\sum_{m=1}^\infty\mathbf{q}_m=1$, and
\begin{align}\mathbf{p}_m(n_g;\rho,\mathbf{q})&\triangleq \frac{\mathbf{q}_m{m\choose n_g}\rho^{n_g}(1-\rho)^{m-n_g}}{\sum_{n=2}^\infty\mathbf{q}_n{n\choose n_g}\rho^{n_g}(1-\rho)^{n-n_g}}.
\end{align}
For the parameters $\mathbf{q}$ and $\rho$ we can use maximum likelihood based on
\begin{align}
\mathbb{P}[n_g|n_g\geq 1]=\frac{\sum_{m=2}^{\infty}\mathbf{q}_m {m\choose n_g}\rho^{n_g}(1-\rho)^{m-n_g}} {\sum_{n=1}^{\infty}\sum_{m=2}^{\infty}\mathbf{q}_m{m\choose n}\rho^n(1-\rho)^{m-n}}.\footnote{To control the number of parameters to be estimated we make the additional restriction $\mathbb{P}[n_{g_0}=n|n_g,\mathbf{x}_g]=\mathbb{P}[n_{g_0}=n|n_g]$ for $n=1,..,\overline{n}$ to derive \eqref{ert}. This contrasts with our identification results, for which this additional restriction is not made.} \label{pok}
\end{align}
In practice we do not sequentially estimate the parameters by maximum likelihood and then by non-linear least squares. Instead, we estimate the parameters jointly by GMM based on the non-linear least squares moment conditions and equality to zero of the expectation of the score of the log-likelihood function based on \eqref{pok}. This framework allows the researcher to use GMM standard errors (as opposed to having to adjust for a two-stage estimator or use a boostrap) and to include additional information when it is available. For example, if the researcher knows which observed groups have missing members (i.e., they observe an indicator for $n_g=n_{g_0}$), they can also use the moment $\mathbb{E}[\mathbf{1}(n_g=n_{g_0})|n_g]=\rho^{n_g}$. Similarly, one can also make a parametric assumption on the distribution of $n_{g_0}$ (i.e., use $\mathbf{q}=\mathbf{q}(\boldsymbol{\lambda})$ for a parameter $\boldsymbol{\lambda}$).

Identically to the empirical likelihood estimator of the probability mass function, the score equations yield $\widehat{\mathbf{q}}_m=0$ for all $m>\widehat{\overline{n}}$, where $\widehat{\overline{n}}$ is the sample maximum of $n_g$, hence the concentrated model is
\begin{align}
\mathbb{E}[\overline{y}_{i}|n_{g},\mathbf{x}_{g}]&=\overline{x}_{i}\sum_{m=2}^{\widehat{\overline{n}}}\mathbf{p}_m(n_g;\rho,\mathbf{q}) \pi(m),\\
\mathbb{P}[n_g|n_g\geq 1]&=\frac{\sum_{m=2}^{\widehat{\overline{n}}}\mathbf{q}_m {m\choose n_g}\rho^{n_g}(1-\rho)^{m-n_g}} {\sum_{n=1}^{\widehat{\overline{n}}}\sum_{m=2}^{\widehat{\overline{n}}}\mathbf{q}_m{m\choose n}\rho^n(1-\rho)^{m-n}}.\label{pok1}
\end{align}
For group uncertainty, we apply non-linear least squares based on \begin{align}\mathbb{E}[\overline{y}_i|n_{g_1},n_{g_2},\mathbf{x}_g]=\overline{x}_i[\psi\pi(n_{g_1})+(1-\psi)\pi(n_{g_2})].\end{align}

\section{Monte-Carlo}\label{mont}

We tailor the design to the well known data on roommates at Dartmouth college studied by \cite{sacerdote01}, \cite{glaeser03} and \cite{angrist14}. The original data comprise 1589 freshman at Dartmouth college who were randomly assigned to dorm-rooms. Fifty three percent of dorm-rooms were doubles, 44 percent were triples and the remaining 3 percent were quads.  The first part of our Monte-Carlo experiment supposes that only a random sample of these students is available. The second part considers the issue of the definition of the relevant peer group, which could be the room or the floor \citep{sacerdote01,glaeser03,angrist14}.

\subsection{Missing data}\label{md}

We consider a design satisfying Assumption \ref{ass2}. The original data comprise a complete sample of freshman (i.e., $\rho=1$). Our design varies $\rho\in\{0.3,0.5,0.7,0.9,1\}$. For $\rho=1$ we set $s_i=1$ for all $i\in\{1,...,N_0\}$ and for $\rho<1$ we set $p_i=\rho+q_i$, where $q_i\overset{i.i.d.}{\sim}\text{Uniform}(-0.1,0.1)$ and $s_i=1$ with probability $p_i$. We set $G_0$ to be the nearest integer to $M/(\rho \mathbb{E}[n_{g_0}])$, where $M$ is discussed below. For each dataset, we draw $G_0$ dorm-rooms from the distribution of dorm-room sizes, which we take to be $n_{g_0}-2\sim {\rm Binomial}(2,0.25)$ so as to match the sample mean and support of the Dartmouth data. 
We use a parametric distribution so as to evaluate the performance of the GMM estimator both with and without a parametric restriction on the group size distribution. 

Since we draw a fixed number of dorm rooms, each with a random size, the number of students $N_0$ varies from one dataset to the next. We then draw an independent random sample of $N$ students, which includes student $i$ with probability $p_i$. Hence $N$ also varies form one dataset to the next. Allowing $G_0$ to depend on $\rho$ as above implies that the size of the observed sample is $N\approx M$ no matter the value of $\rho$. Setting $M=1600$, our design matches the sample size of the Dartmouth data. We thus consider the performance of our method had the Dartmouth data been obtained from a random sample of freshman, rather than all freshman, holding the number of observations constant as we vary $\rho$. We also consider $M=8000$ to study the performance of the method under different sample sizes.

\cite{sacerdote01} exploits random assignment of freshmen to dorms in the Dartmouth data to identify peer effects in educational attainment, measured by freshman GPA. It is argued that endogenous effects are difficult to identify due to the reflection problem (see \cite{manski93}). 
\cite{sacerdote01} focuses instead on credibly identifying the reduced form effect of room-mate high school attainment on freshman GPA (see column 5 of Table 3 in \cite{sacerdote01}). The estimated peer effect is positive and statistically significant, though smaller in magnitude than own high-school attainment. This reduced form regression has $R^2=0.19$. 

To match this setting as well as possible, in our design we set the own effect to be $\gamma=1$ and the endogenous effect to be $\beta=0$, hence the reduced form effect of room-mate's high-school attainment is equal to the contextual effect, which we set to be $\delta=0.5$. We set $x_i\overset{i.i.d.}{\sim}\mathcal{N}(0,1)$, $\epsilon_i\overset{i.i.d.}{\sim}\mathcal{N}(0,\sigma^2)$ and $\alpha_{g_0}\overset{i.i.d.}{\sim}\mathcal{N}(1,\sigma^2)$ and choose $\sigma^2=2(\gamma^2+\delta^2/\mathbb{E}[n_{g_0}-1])$. For the expected dorm-room size, this choice of $\sigma^2$ corresponds to fraction 0.8 of $\mathbb{V}[y_i]$ being due to $\alpha_{g_0(i)}+\epsilon_i$ and 0.2 being due to $x_i\gamma+(1-n_{g_0})^{-1}\sum_{j:g_0(j)=g_0(i),j\neq i}x_j$, hence a population $R^2$ of 0.2 in a reduced form regression without fixed effects, as estimated by \cite{sacerdote01}.

\subsubsection{Results}

Table \ref{resultsmc1} presents the results. We compare four models; one in which the observed groups are treated as if they are the groups (`MS', estimated by NLS), one in which the group size is known (`K', estimated by NLS), one in which the group size is unknown (`U', estimated by GMM) and a modification of U in which the parametric assumption $n_{g_0}-2\sim {\rm Binomial}(2,\omega)$ is additionally imposed (`U-P', estimated by GMM). Model MS is miss-specified when $\rho<1$. The other models are correctly specified but use information and assumptions in different ways. Model (K) is the full information benchmark with which we compare models (U) and (U-P). Models (U) and (U-P) take the upper bound on the support of the group sizes to be known as $\overline{n}=4$, which, in the context of the Dartmouth data implies that the researcher knows that there are no rooms of more than four students. 

Beginning with the contextual effects only specification ($\beta=0$ is imposed) and $\rho=1$, we see that the results are very similar (identical to three decimal places) for all four models. As $\rho$ decreases, the distributions of the NLS estimator of $\delta$ and $\gamma$ of model (MS) shift closer to zero. For $\rho\leq0.5$, the bias becomes sizeable. In contrast, the estimators for the other models remain approximately unbiased for all values of $\rho$, though their root mean squared error (RMSE) increases as $\rho$ decreases. This is at least partly because, as $\rho$ decreases the number of groups with only one observed member increases, and these groups provide no information on the parameters due to the within-specified-group transformation (i.e., because $\overline{y}_i=\overline{x}_i=0$ when $n_g=1$).

The NLS estimator for model (K) performs at least as well as the GMM estimator for model (U) in terms of RMSE. This is unsurprising since it requires known group sizes in order to be implemented. For $\rho=0.9$ the difference in RMSE is small (0.086 vs 0.099 for $\delta$ when $M=8000$), though the gap widens as $\rho$ decreases. This is likely because, as $\rho$ decreases, there is less variation in the specified group size $n_g$, and hence, for a given specified group size, more uncertainty as to the group size $n_{g_0}$. The additional parametric assumption on the group size distribution used in model (U-P) makes little difference when $\rho\geq 0.7$, but for smaller values the difference in RMSE between the GMM estimators of models (U) and (U-P) becomes non-neglible.

Moving on to the specification with contextual and endogenous effects, it is clear that, though identified due to there being at least three group sizes, there is insufficient group size variation in this design to reliably separate the two, as suggested by \citep{sacerdote01}. This is likely because only around 6\% of rooms were quads. In particular, the RMSE on $\beta$ is large and all estimators of $\beta$ are biased upwards, whereas estimators of $\delta$ tend to be biased downwards. Nevertheless, the qualiative conclusions regarding the relative performance of the estimators of the four models are the same as for the specifications with contextual effects only.

\subsection{Group uncertainty}

\cite{sacerdote01}, \cite{glaeser03} and \cite{angrist14} all discuss the appropriate definition of the peer group, which may either be the room, the floor or the entire dorm. 
We focus here on the distinction between the room and the floor, and consider a design in which Assumption \ref{ass3} holds. To match the Dartmouth data, we first draw the rooms as described above. We then draw $f_1\in\{1,2,...,5\}$ uniformly, and take rooms $\{1,2,...,f_1\}$ to comprise the first floor. We then draw another integer $f_2\in\{1,2,...,5\}$ and take rooms $\{f_1+1,...,f_1+f_2\}$ to comprise the second floor. We proceed in this way until every room has been allocated to a floor. In the Dartmouth data, the sample mean number of students per floor is close to 8. This data generating process yields an expected number of students per floor of 7.5. All other aspects of the data generating process remain unchanged and we vary $\psi\in\{0.2,0.4,0.6,0.8\}$.

\subsubsection{Results}

Table \ref{resultsmc2} presents the results. We compare four models; one in which the specified group is the room (`R'), one in which the specified group is the floor (`F'), one in which the researcher knows whether the group is the room or the floor (`K') and one in which the group is unknown (`U'). All are estimated by NLS. Models (R) and (F) are miss-specified because have $0<\psi<1$ in all designs. The other models are correctly specified but use information and assumptions in different ways. Model (K) is the full information benchmark with which we compare model (U).

For brevity, the discussion focuses on the contextual effects only specifications. Specifying the group to be the room (model (R)) leads to downwards bias of $\delta$. The bias and RMSE grow as $\psi$ decreases. This is because the proportion of incorrectly specified groups grows as $\psi$ decreases. For the same reason the RMSE from specifying the group to be the floor (model (F)) increases as $\psi$ increases. 

The results suggest that the estimator of model (F) is unbiased even when $\psi>0$. This is coincidental. It just so happens that in this design the three sources of bias (the terms $a,b$ and $c$ discussed in Example 1 in Section \ref{msg}) offset one another. To demonstrate this, Table \ref{resultsmc3} modifies the design by instead setting $\alpha_{g_0(i)}\overset{i.i.d.}{\sim}\mathcal{N}(n_{g_0}^-1\sum_{j:g_0(j)=g_0(i)}x_j,\sigma^2)$. This has no impact on the estimators of models (R), (K) and (U) because all are based on room fixed effects.\footnote{For these estimators the results are numerically identical to results reported in Table \ref{resultsmc2} because identical seeds were used to draw the data for each replication.} However, it introduces substantial bias for the estimator of model (F), which uses floor fixed effects.

Comparing the results for models (K) and (U), as expected the RMSE of the latter is larger since it does not require knowledge of the groups. Nevertheless, depending on the sample size and value of $\psi$, the NLS estimator of model (U) has small bias and sufficiently small RMSE so as to be distinguishable from zero. Considering now the specifications with endogenous and contextual effects, though weakly identified, the main qualiative conclusions regarding the relative performance of the estimators of the four models are the same as for the specifications with contextual effects only. Notice also that once endogenous effects are introduced the estimator of model (F) often has larger bias and RMSE than that of model (U). This is particularly true for $\delta$.\\

\section{Application}

In this section, we use unique employer-employee matched data to study how lawyers' quit-to-exit responses depend on their own ability and that of their peers. The dataset is collected from the Shanghai Bar Association and includes a full history of law firm composition and lawyers' career changes from 2009-2016. Exploiting lawyers' annual registration records, we observe when they cancel legal practice in Shanghai, which we refer to as their quit-to-exit. On quitting, lawyers may change occupations, practice law in other regions, or return to education. Since legal practice is among the highest-income occupations and Shanghai is one of China's highest-paying cities, we argue that quit-to-exit in this context is likely a result of lesser persistence in a highly competitive, incentivized environment, similar to exits in the education and labor literature \citep{thiemann2022persistent, wasserman2018gender}. 
  
Our analysis focuses on lawyers' quit-to-exit in 2016 and how it is affected by their own ability and that of their peers. Throughout our analysis, we focus on associate lawyers, excluding partners and directors. This is because associates and partners/directors are unlikely to perceive one another as direct competitors. The behavior, ability and characteristics of partners and directors in the firm is accounted for by firm fixed effects. From this point onwards, we refer to associate lawyers simply as lawyers. 

To measure lawyers' ability, we merge the registration records with Shanghai court judgements. In this dataset, lawyers' performance can be more reliably measured for civil litigation, so we focus on law firms and lawyers whose main business is in civil law. Our analysis is thus restricted to a sample of law firms in which most lawyers complete at least one civil case annually and lawyers who do not declare criminal law as one of the three specialized fields.\footnote{The performance in criminal law is not well measured because: 1) We only observe a small share of criminal judgements and many are restricted for confidentiality reason; 2) For criminal cases, the case fees are identical within the same case category (e.g. theft or homicide), so we cannot use case fees to measure case size as what we do for civil cases. Nevertheless, our results are robust to use number of civil and criminal cases to create the ability measure and run the analysis for all (civil and criminal) lawyers, and are robust to including all law firms in the analysis (see Table \ref{tab:approb}).} To measure performance in civil litigation, we use lawyers' fee-weighted caseload in the previous three years (2013-2015). Though we observe case outcomes (win or lose), we do not use this to measure ability because it is likely that high-ability lawyers take on more difficult cases. 

To compute our ability measure, we perform the following steps: 1) We extract the case fee (measured in thousand yuan) for each civil case, which is the amount paid to file the case to the court and is calculated based on the disputed amount of the case; 2) We divide the case fee by the number of lawyers representing the case, according to the lawyers being listed in the court judgement; and 3) We repeat the previous two steps for each civil case and sum all fee-weighted cases undertaken from 2013-2015, and 4) We compute the annual average caseload by dividing by the number of years of practice between 2013 and 2015.\footnote{For lawyers who started practice after 2013, the annual caseload in 2013 (and 2014 if started after 2014) would be zero. To include these lawyers and make them comparable, we use the annual average caseload, rather than the three-year sum, by excluding the year(s) before the lawyer's legal practice date.} 

The fee-weighted annual average caseload is a proxy for ability because it reflects past performance and income generated, hence can be used to predict one's career prospects. Generally, lawyers in China charge by case, rather than by work hours \citep{liu2006client}. For civil litigation, lawyers' fees are typically charged based on the disputed amount \citep{michelson2006practice}, thus high-ability lawyers are more likely to work on high fee cases, regardless of whether cases are assigned by partners or obtained by the lawyer themselves. Nevertheless, we are aware that non-litigation business is not measured (e.g., Initial Public Offerings and Mergers/Acquisitions). To this end, we replicate our analysis in small- to medium-size firms because major non-litigation business is concentrated in large firms.

The effect of interest is that of peer ability on quit-to-exit. To measure peer ability we use both the peer average fee-weighted caseload and the proportion of high-ability peers, where high-ability is defined as having a fee-weighted caseload in the top quartile (i.e., $>24.44$ in our data). We believe that the latter better captures the competitive environment through which lawyers compete for promotion to a small number of senior positions (team leaders, partners and directors). 

We consider two group structures through which peer effects may operate. Group F supposes that peer effects operate among all lawyers working at the same law firm. Alternatively, we might expect stronger peer effects among similar-age lawyers, who are likely to be at a similar career stage. To account for this possibility, we split lawyers into two cohorts with a cut-off age of 35  (inclusive) for the younger cohort. Group F-C supposes that peer effects only operate through lawyers in the same cohort at the same law firm. 

Our sample includes 8,448 lawyers working at 755 law firms during 2016. Table \ref{tab:sum} summarizes the data. As shown in the left panel, 3.6\% of lawyers quit-to-exit, 45\% are female, 38\% have a graduate degree (Masters or PhD), the average age is around 36, average experience is around 8 years, and the average fee-weighted annual caseload is 20 thousand yuan. In the right panel of Table \ref{tab:sum}, we present summary statistics for the 732 small- to medium-size firms, which are largely similar to the sample of all firms.

We estimate \eqref{struc} taking the dependent variable to be an indicator for quit-to-exit in 2016. Even though the dependent variable is binary, the linear model is well defined from both an econometric and economic perspective, and has desirable computational and statistical properties, especially when there are group fixed effects (see \cite{bou20}). 

Individual covariates comprise an indicator for being female, an indicator for having a graduate degree, age, years of experience, quadratic terms in age and experience, and fee-weighted caseload. We consider contextual effects only (i.e., $\beta=0$ is imposed) because we do not expect strategic interactions in quit-to-exit behavior. To account for within-firm dependence, we cluster standard errors by firm. All soecifications include either firm or firm-cohort fixed effects.

\subsection{Baseline results}

Our baseline results are reported in Table \ref{tab:app}. Models (1)-(2) respectively use group structures F and F-C. We find a positive yet statistically insignificant peer effect in fee-weighted caseload. In group F, the magnitude of the peer effect is around one third of the effect of a lawyer's own fee-weighted caseload, which is negative and statistically significant at the 0.01 level. For group F-C, the magnitude of the peer effect is similar to that of own fee-weighted caseload. Models (3)-(4) consider instead the proportion of high-ability peers in groups F and F-C respectively, for which we find larger peer effects. The coefficient of 0.0544 in model (3) is statistically significant at the 0.05 level and indicates that a one standard deviation (17.2 pp) increase in the proportion of high-ability peers leads to 26.0 percent (0.9 pp) increase in lawyers' quit-to-exit probability. We find a similar effect in model (4). In both models we find a notable gender gap, in that female lawyers are around 35.8-41.1 percent (1.3-1.5 pp) more likely to quit than male lawyers. This is consistent with previous findings that women are more likely to discontinue in competitive environments \citep{hunt2016women, wasserman2018gender}. We also find evidence that quits are negatively but concavely correlated with own age and negatively correlated with own fee-weighted caseload, which is consistent with \cite{jovanovic1979job}. 

Next, in models (5)-(6) we estimate the impact of peer ability separately for those above and below 35. We find that the peer effect is stronger in the younger cohort. The coefficient of 0.0767 in model (5) is statistically significant at the 0.01 level and indicates that a one standard deviation (19.3 pp) increase in the proportion of high-ability peers leads to 35.9 percent (1.48 pp) increase in lawyers' quit-to-exit likelihood. In contrast, we find a smaller effect among those older than 35, which is not statistically distinguishable from zero. We also find a larger gender gap among those below 35. Overall, we find stronger quit-to-exit effects among those under 35, which is consistent with \cite{jovanovic1979job}.

In Table \ref{tab:appnl50}, we restrict our analysis to small- to medium- size firms to alleviate the concern that our ability measure cannot capture non-litigation performance. We limit the sample of firms to those have 50 or fewer lawyers (i.e. 732 out of 755 firms). Our results are qualitatively identical the baseline specification. Another concern is on lawyers close to the retirement age, who may intentionally reduce caseload before the retirement. To that end, we replicate our preferred models (models (3)-(4) in Table \ref{tab:app}) and we exclude lawyers beyond age 55, the age at which most employees in China are eligible to claim their pension. As shown in models (1)-(2) of Table \ref{tab:approb}, we still find similar adverse peer ability effects, indicating that retiring lawyers are not driving the estimated effects. Next, in models (3)-(4), we conduct another robustness check by including all law firms in Shanghai. The similar effects we find in these two columns suggest that caseload broadly serves as a reliable ability proxy in the legal profession context. Overall, the similar effects found in alternative samples in models (1)-(4) and in Table \ref{tab:appnl50} imply that the estimated peer ability effects generally exist in our context. Lastly, in models (5)-(6), we consider an alternative definition using the number of high-ability peers in the specific groups F and F-C. The results are also in line with our baseline findings.

In Table \ref{tab:apphet}, we conduct a heterogeneity analysis. In models (1)-(2), we examine the impact of high-ability peers on lawyers whose fee-weighted caseload is below median. Consistent with \cite{antecol2016peer} and \cite{booij2017ability}, these low- to medium-ability individuals drive the negative peer ability effects. In models (3)-(4), we study whether women are more sensitive to peer ability effects. Despite the significant gender gap of persistence, we find no evidence suggesting that women are more sensitive to peer ability effects. In models (5)-(6), we check whether the magnitude of peer ability effects vary by firm size by interacting the proportion of high-ability peers with an indicator for firms below the median size. The coefficient on the interaction term suggests that lawyers in small firms are not additively more prone to peer ability effects.

\subsection{Missing data}

\begin{figure}[H]
	\centering
	    \caption{Peer effects in lawyers' quit-to-exit: Distribution of number of associate lawyers ($n_{g_0}$)}
    \label{fig:app0}
    \includegraphics[width=0.9\textwidth]{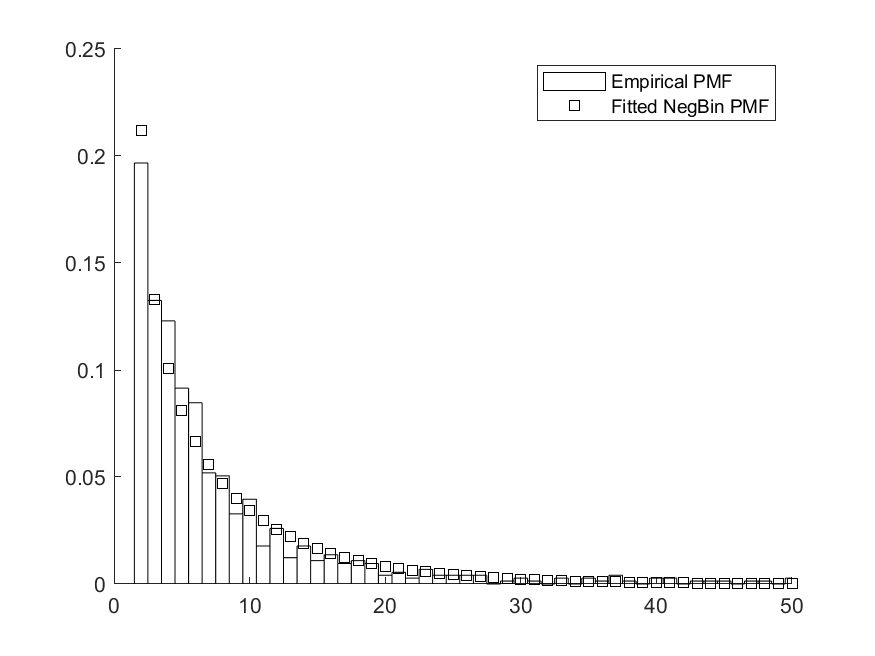}
 \begin{flushleft}
        \footnotesize{\textbf{Notes:} The minimum of $n_{g_0}$ in our sample is 2. The negative binomial distribution is thus fitted using $n_{g_0}-2$.}
\end{flushleft}
\end{figure}

We now ask whether we could have obtained similar findings had only a random sample of lawyers with firm identifiers been available. To do this, we draw random 1000 random sub-samples of lawyers using the sampling process from our Monte-Carlo experiment (see Section \ref{md}). We focus on the baseline specification for group F (model (3) of Table \ref{tab:app}), consider 50\% and 70\% subsamples (i.e., $\rho=0.5$ and $\rho=0.7$). The minimum value of $n_{g_0}$ in our sample is 2, which we use as a lower bound for the estimator with unknown group sizes. We make a parametric restriction on the distribution of $n_{g_0}-2$, supposing that it follows a negative binomial distribution. Figure \ref{fig:app0} depicts the empirical distribution for the original sample and a fitted negative binomial distribution, which provides a good approximation.

\begin{figure}[H]
	\centering
	\caption{Peer effects in lawyers' quit-to-exit: Random  50\% subsamples}    \label{fig:app50}
    \includegraphics[height=21cm]{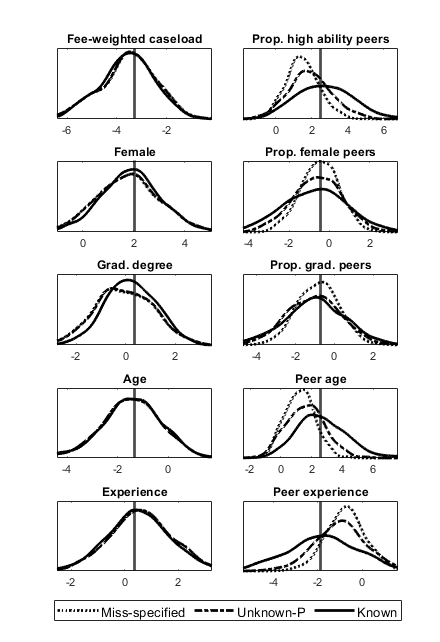}
 \begin{flushleft}
        \footnotesize{\textbf{Notes:} We report the distribution (fitted by kernel density estimation) of the following estimators based on 1000 random subsamples of proportion $\rho=0.5$ of lawyers. `Miss-specified' refers to the NLS estimator obtained by treating the observed firms as if they were the firms (i.e., based on the moment
$\mathbb{E}[\overline{y}_i|n_g,\mathbf{x}_g]=\overline{x}_i\pi(n_g)$). `Known' is the NLS estimator under Assumption \ref{ass1} (i.e., based on the moment
$\mathbb{E}[\overline{y}_i|n_{g_0},\mathbf{x}_g]=\overline{x}_i\pi(n_{g_0})$). `Unknown-P' is the GMM estimator under Assumption \ref{ass2} imposing the parametric restriction $n_{g_0}-2\sim {\rm NegativeBinomial}(m,p)$, where both parameters are estimated. The distributions are rescaled by the standard error obtained from full sample estimation. Vertical lines correspond to the rescaled full sample estimate (i.e., the t-statistic for the test of statistical significance). See model (3) of Table \ref{tab:app} for full sample estimates.}
\end{flushleft}
\end{figure}

\begin{figure}[H]
	\centering
	\caption{Peer effects in lawyers' quit-to-exit: Random 70\% subsamples}    \label{fig:app70}
    \includegraphics[height=21cm]{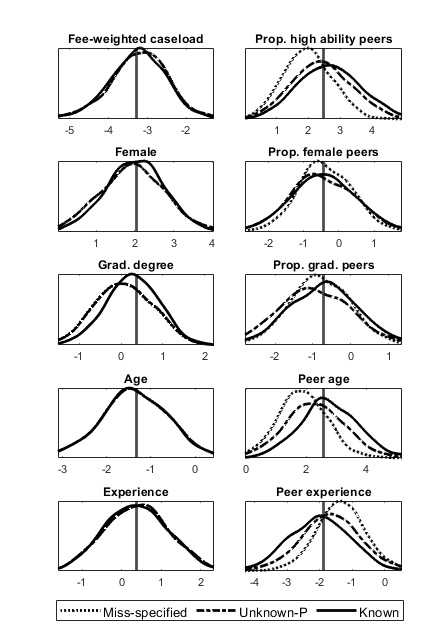}
 \begin{flushleft}
        \footnotesize{\textbf{Notes:} We report the distribution (fitted by kernel density estimation) of the following estimators based on 1000 random subsamples of proportion $\rho=0.7$ of lawyers. `Miss-specified' refers to the NLS estimator obtained by treating the observed firms as if they were the firms (i.e., based on the moment
$\mathbb{E}[\overline{y}_i|n_g,\mathbf{x}_g]=\overline{x}_i\pi(n_g)$). `Known' is the NLS estimator under Assumption \ref{ass1} (i.e., based on the moment
$\mathbb{E}[\overline{y}_i|n_{g_0},\mathbf{x}_g]=\overline{x}_i\pi(n_{g_0})$). `Unknown-P' is the GMM estimator under Assumption \ref{ass2} imposing the parametric restriction $n_{g_0}-2\sim {\rm NegativeBinomial}(m,p)$, where both parameters are estimated. The distributions are rescaled by the standard error obtained from full sample estimation. Vertical lines correspond to the rescaled full sample estimate (i.e., the t-statistic for the test of statistical significance). See model (3) of Table \ref{tab:app} for full sample estimates.}
\end{flushleft}
\end{figure}

The results are reported in Figures \ref{fig:app50} ($\rho=0.5$) and \ref{fig:app70} ($\rho=0.7$). As in our Monte-Carlo experiment, we consider the estimator based on miss-specified peer group sizes (i.e., incorrectly supposing that $n_g=n_{g_0}$), known sizes and unknown sizes. Looking at the left hand columns we see that miss-specification has little consequence for estimates of the effect of a lawyer's own characteristics. However, looking at the right hand columns we see that miss-specification biases estimates of peer effects towards zero. Comparing Figure \ref{fig:app50} with Figure \ref{fig:app70}, we see that the bias in in the peer effects is larger for $\rho=0.5$ than for $\rho=0.7$. Unsurprisingly, the variability of the estimators is also larger for $\rho=0.5$ than for $\rho=0.7$. These findings are qualitatively identical to those of our Monte-Carlo experiment, but use real data and larger group sizes.

\subsection{Group uncertainty}

Model (7) of Table \ref{tab:app} allows for uncertainty as to whether F or F-C is the relevant peer group. The point estimate of $\psi$ (i.e., the probability of F-C) is 0.562, which suggests considerable heterogeneity, though it is not precisely estimated. Despite this, we find similar peer effects to the models based on groups F and F-C. We obtain similar results in the sample of small-medium firms (see Table \ref{tab:appnl50}).

\section{Conclusion}\label{disc}

Our identification results and empirical work demonstrate that it is possible to conduct empirical analysis of peer effects despite missing data and group uncertainty. Regarding missing data, we require only that the researcher has access to a sample of individuals with outcomes, exogenous characteristics and group identifiers. We propose a method which does not require information on group size, nor on individuals which are not sampled, nor on whether such individuals even exist. In principle, and subject to the limitations discussed below, this opens up the possibility of peer effects studies based on widely available individual level survey data (provided that it contains group identifiers). 
We also show that peer effects can be identified under group uncertainty. Future work may extend our results to incorporate both missing individuals and group uncertainty simultaneously. 

A limitation of our work is that our results are specific to the group interactions model we consider. Under more general network structures (e.g., social networks), it is not the case that the specified group fixed effects reduce the problem of miss-specification to the problem of inferring the group size. Future work may seek to bridge this divide. 

\singlespacing
\bibliographystyle{ier}
\bibliography{refs}
\onehalfspacing
\section*{Appendix}

\begin{proof}of Proposition \ref{the1}\\

\noindent The first result was established by \cite{davezies09}. \\

\noindent To establish the second result, first note that Assumption \ref{ass2} implies $\mathbb{E}[\overline{u}_i|n_{g},\mathbf{x}_{g}]=0$ for all $i\in\mathcal{S}$. Let us now fix $\mathbf{x}_g=\mathbf{x}$ for some $\mathbf{x}$ in its support. We begin by identifying $\rho$ and the distribution of $n_{g_0}|\mathbf{x}_g=\mathbf{x}$ using the observable distribution of $n_{g}|n_{g}\geq 1,\mathbf{x}_g=\mathbf{x}$ and the fact that Assumption \ref{ass2} implies $$n_{g}|n_{g_0},\mathbf{x}_g=\mathbf{x}\sim \text{Binomial}(n_{g_0},\rho)$$ 
By the boundedness assumption, there exists $\overline{n}<\infty:\mathbb{P}[n_{g_0}=\overline{n}]>0,\mathbb{P}[n_{g_0}\leq \overline{n}]=1$. Note that, because $\rho>0$, $\overline{n}$ is identifiable as the integer $n$ such that $\mathbb{P}[n_{g}=n|n_{g}\geq 1]>0$. This implies that there exists identifiable $\overline{n}(\mathbf{x})<\infty:\mathbb{P}[n_{g_0}=\overline{n}(\mathbf{x})|\mathbf{x}_g=\mathbf{x}]>0,\mathbb{P}[n_{g_0}(\mathbf{x})\leq \overline{n}(\mathbf{x})|\mathbf{x}_g=\mathbf{x}]=1$.

Let $\mathbf{p}_n(\mathbf{x})=\mathbb{P}[n_{g}=n|n_{g}\geq 1, \mathbf{x}_g=\mathbf{x}]$ and $\mathbf{q}_m(\mathbf{x})=\mathbb{P}[n_{g_0}=m|\mathbf{x}_g=\mathbf{x}]$. Under Assumption \ref{ass2}, $\mathbf{p}(\mathbf{x})=(\mathbf{p}_n(\mathbf{x}))_{n=1,...,\overline{n}(\mathbf{x})}$ and $\mathbf{q}(\mathbf{x})=(\mathbf{q}_m(\mathbf{x}))_{m=1,...,\overline{n}(\mathbf{x})}$ verify
\begin{align}
\mathbf{p}(\mathbf{x})/\mathbf{p}_{\overline{n}(\mathbf{x})}(\mathbf{x})=\mathbf{A}(\mathbf{x},\rho)\mathbf{s}(\mathbf{x})\rho^{-\overline{n}(\mathbf{x})}\label{p1}
\end{align}
where $\mathbf{A}(\mathbf{x},\rho)=(\mathbf{A}_{ij}(\mathbf{x},\rho))_{i=1,...,\overline{n}(\mathbf{x}),j=1,...,\overline{n}(\mathbf{x})}$ is an upper triangular Binomial matrix with entries, 
$$\mathbf{A}_{ij}(\mathbf{x},\rho)=\begin{cases}{j\choose i}\rho^i(1-\rho)^{j-i}&i\leq j\\0&\text{otherwise} \end{cases}$$ and $\mathbf{s}(\mathbf{x})=\mathbf{q}(\mathbf{x})/\mathbf{q}_{\overline{n}(\mathbf{x})}(\mathbf{x})$. By construction, $\mathbf{s}_{\overline{n}(\mathbf{x})}(\mathbf{x})=1$ and $\sum_{m=1}^{\overline{n}(\mathbf{x})}\mathbf{s}_m(\mathbf{x})=1/\mathbf{q}_{\overline{n}(\mathbf{x})}(\mathbf{x})$. We also know $\mathbf{q}_1(\mathbf{x})=0$ because $n_{g_0}\geq 2$, hence we know $\mathbf{s}_1(\mathbf{x})=0$. The remaining entries of $\mathbf{s}(\mathbf{x})$ are non-negative but unknown.

Now suppose that there exists $\widetilde{\rho}$ and $\widetilde{\mathbf{s}}(\mathbf{x})$ which also verify \eqref{p1}. Then we have
\begin{align}
\mathbf{s}(\mathbf{x})&=\left(\frac{\rho}{\widetilde{\rho}}\right)^{\overline{n}(\mathbf{x})}\mathbf{A}(\mathbf{x},\rho)^{-1}\mathbf{A}(\mathbf{x},\widetilde\rho)\widetilde{\mathbf{s}}(\mathbf{x})\\
&=\mathbf{B}(\mathbf{x},\rho,\widetilde{\rho})\widetilde{\mathbf{s}}(\mathbf{x})\label{p2}
\end{align}
where $\mathbf{A}(\mathbf{x},\rho)^{-1}$ exists since $|\mathbf{A}(\mathbf{x},\rho)|=\rho^{T(\overline{n}(\mathbf{x}))}$ where $T(\overline{n}(\mathbf{x}))$ is the $\overline{n}(\mathbf{x})^{th}$ triangular number and by assumption $\rho>0$. The matrix $\mathbf{B}(\mathbf{x},\rho,\widetilde\rho)=(\mathbf{B}_{ij}(\mathbf{x},\rho,\widetilde\rho))_{i=1,...,\overline{n}(\mathbf{x}),j=1,...,\overline{n}(\mathbf{x})}$ is an upper triangular matrix with entries,
$$\mathbf{B}_{ij}(\mathbf{x},\rho,\widetilde\rho)=\begin{cases}\left(\frac{\rho}{\widetilde{\rho}}\right)^{\overline{n}(\mathbf{x})}{j\choose i}\widetilde{\rho}^i\rho^{-j}(\rho-\widetilde\rho)^{j-i}   &i\leq j\\0&\text{otherwise} \end{cases}.$$
The final equation in the system in \eqref{p2} is redundant because we already established that $\mathbf{s}_{\overline{n}(\mathbf{x})}=\widetilde{\mathbf{s}}_{\overline{n}(\mathbf{x})}=1$ by construction. Rearranging the penultimate equation yields
\begin{align}
\rho=\widetilde{\rho}\left(\frac{\mathbf{s}_{\overline{n}(\mathbf{x})-1}(\mathbf{x}) +\overline{n}(\mathbf{x})}{\widetilde{\mathbf{s}}_{\overline{n}(\mathbf{x})-1}(\mathbf{x}) +\overline{n}(\mathbf{x})}\right)\label{pr}
\end{align}
Denoting the term in parentheses by $\Delta(\mathbf{x})$, injecting \eqref{pr} into the definition of $\mathbf{B}(\mathbf{x},\rho,\widetilde\rho)$ yields $\mathbf{B}(\mathbf{x},\rho,\widetilde\rho)=\mathbf{C}(\Delta(\mathbf{x}))$ with entries
$$\mathbf{C}_{ij}(\Delta(\mathbf{x}))=\begin{cases}{j\choose i}\Delta(\mathbf{x})^{\overline{n}(\mathbf{x})-j} (\Delta(\mathbf{x})-1)^{j-i} &i\leq j\\0&\text{otherwise} \end{cases},$$
hence we can re-write the system in \eqref{p2} as
\begin{align}
\mathbf{s}(\mathbf{x})&=\mathbf{C}(\Delta(\mathbf{x}))\widetilde{\mathbf{s}}(\mathbf{x})\label{p3}.
\end{align}
From its definition, it is clear that $\Delta(\mathbf{x})>0$ and when $\Delta(\mathbf{x})=1$ we have $\rho=\widetilde\rho$ (due to \eqref{pr}), $\mathbf{s}(\mathbf{x})=\widetilde{\mathbf{s}}(\mathbf{x})$ (due to \eqref{p2}) and hence $\mathbf{q}(\mathbf{x})=\widetilde{\mathbf{q}}(\mathbf{x})$ (because $\sum_{m=1}^{\overline{n}(\mathbf{x})}\mathbf{s}_m(\mathbf{x})=1/\mathbf{q}_{\overline{n}(\mathbf{x})}(\mathbf{x})$). We now show that $\Delta(\mathbf{x})=1$ by ruling out $\Delta(\mathbf{x})>1$ and $\Delta(\mathbf{x})<1$ by contradiction.

Suppose first that $\Delta(\mathbf{x})>1$. Then, since $\mathbf{s}_1(\mathbf{x})=0$, the first equation in \eqref{p3} is
$$
0=\sum_{j=1}^{\overline{n}(\mathbf{x})}j\Delta(\mathbf{x})^{\overline{n}(\mathbf{x})-j} (\Delta(\mathbf{x})-1)^{j-1}\widetilde{\mathbf{s}}_j(\mathbf{x})
$$
The right hand side is strictly positive because by construction at least one entry of $\widetilde{\mathbf{s}}(\mathbf{x})$ must be strictly positive. Hence we cannot have $\Delta(\mathbf{x})>1$.

Suppose instead that $\Delta(\mathbf{x})<1$. Using the first $\overline{n}(\mathbf{x})-2$ equations in \eqref{p3} to solve for the $\overline{n}(\mathbf{x})-2$ unknowns $\widetilde{\mathbf{s}}_2(\mathbf{x}),\widetilde{\mathbf{s}}_3(\mathbf{x}),...,\widetilde{\mathbf{s}}_{\overline{n}(\mathbf{x})-1}(\mathbf{x})$, we obtain
\begin{align}
\begin{pmatrix}
\widetilde{\mathbf{s}}_2(\mathbf{x})\\
\widetilde{\mathbf{s}}_3(\mathbf{x})\\
\vdots\\
\widetilde{\mathbf{s}}_{\overline{n}(\mathbf{x})-1}(\mathbf{x})
\end{pmatrix}
=\mathbf{D}(\Delta(\mathbf{x}))^{-1}\left[
\begin{pmatrix}
0\\
\mathbf{s}_2(\mathbf{x})\\
\vdots\\
\mathbf{s}_{\overline{n}(\mathbf{x})-2}(\mathbf{x})
\end{pmatrix}
-\mathbf{d}(\Delta(\mathbf{x}))
\right],\label{p4}
\end{align}
where $\mathbf{D}(\Delta(\mathbf{x}))$ is the sub-matrix of $\mathbf{C}(\Delta(\mathbf{x}))$ formed from rows $1,2,...,\overline{n}(\mathbf{x})-2$ and columns $2,3,...,\overline{n}(\mathbf{x})-1$ and $\mathbf{d}(\Delta(\mathbf{x}))$ is the sub-matrix of $\mathbf{C}(\Delta(\mathbf{x}))$ formed from rows $1,2,...,\overline{n}(\mathbf{x})-2$ and column $\overline{n}(\mathbf{x})$. Injecting \eqref{p4} into the right hand side of the penultimate equation of \eqref{p3} yields

$$
\mathbf{s}_{\overline{n}(\mathbf{x})-1}(\mathbf{x})=\begin{cases}-\frac{\overline{n}(\mathbf{x})(1-\Delta(\mathbf{x}))}{\overline{n}(\mathbf{x})-1}&\overline{n}(\mathbf{x})=2 \\-\frac{\sum_{j=2}^{\overline{n}(\mathbf{x})-2}j(1-\Delta(\mathbf{x}))^{j-2} \mathbf{s}_{j}(\mathbf{x})}{(\overline{n}(\mathbf{x})-1)(1-\Delta(\mathbf{x}))^{\overline{n}(\mathbf{x})-3}}-\frac{\overline{n}(\mathbf{x})(1-\Delta(\mathbf{x}))}{\overline{n}(\mathbf{x})-1}&\overline{n}(\mathbf{x})\geq 3\end{cases}
$$
Since $\Delta(\mathbf{x})\in(0,1)$ and the entries of $\mathbf{s}(\mathbf{x})$ are non-negative, the right hand side is strictly negative whilst the left hand side is non-negative. Hence we cannot have $\Delta(\mathbf{x})<1$. So $\Delta(\mathbf{x})=1$ and $\rho$ and $\mathbf{q}(\mathbf{x})$ are identified.

Provided that there is variation in the elements of $\mathbf{x}$, the conditional moment
\begin{align}
\mathbb{E}[\overline{y}_{i}|n_{g}=n,\mathbf{x}_{g}=\mathbf{x}]&= \overline{x}_{i}\varphi(n,\mathbf{x})\label{redEE}
\end{align}
identifies $\varphi(n,\mathbf{x})$ for all such $(\mathbf{x},n\geq 2)$ in the support of $(\mathbf{x}_g,n_g)$ (it is not identified when $n=1$ nor when $\mathbf{x}=c\iota_n$ for some constant $c$ because in these cases $\overline{y}_{i}=\overline{x}_{i}=0$). The rest of the proof restricts attention to the case in which there is variation in the elements of $\mathbf{x}$.

Define $\mathbf{r}_n(\mathbf{x})=\mathbb{P}[n_{g}=n|\mathbf{x}_g=\mathbf{x}]$ and $\mathbf{r}(\mathbf{x})=(\mathbf{r}_n(\mathbf{x}))_{n=1,...,\overline{n}(\mathbf{x})}$. Since $\mathbf{q}(\mathbf{x})$ and $\rho$ are identified, $\mathbf{r}(\mathbf{x})$ is also identified
using
$$
\mathbf{r}(\mathbf{x})=(\iota_{\overline{n}(\mathbf{x})}'\mathbf{A}(\mathbf{x},\rho)\mathbf{q}(\mathbf{x}))\mathbf{p}(\mathbf{x}).
$$
Now denote $\mathcal{Q}(\mathbf{x})=\{m\in\{2,...,\overline{n}(\mathbf{x})\}: \mathbf{q}_m(\mathbf{x})>0\}$, the largest element of which is $\overline{n}(\mathbf{x})$. Moreover, since $\rho>0$, the support of $\mathbf{r}(\mathbf{x})$ is $\mathcal{R}(\mathbf{x})=\{1,...,\overline{n}(\mathbf{x})\}$ if $\rho\in(0,1)$ and $\mathcal{R}(\mathbf{x})=\mathcal{Q}(\mathbf{x})$ if $\rho=1$. Let $\widehat{\boldsymbol{\pi}}(\mathbf{x})=(\pi(m))_{m\in \mathcal{Q}(\mathbf{x})}$, $\widehat{\mathbf{q}}(\mathbf{x})=(\mathbf{q}_m(\mathbf{x}))_{m\in \mathcal{Q}(\mathbf{x})}$, $\widehat{\mathbf{r}}(\mathbf{x})=(\mathbf{r}_n(\mathbf{x}))_{n\in \mathcal{R}(\mathbf{x})\backslash 1}$ and $\widehat{\mathbf{A}}(\mathbf{x},\rho)$ be the sub-matrix of $\mathbf{A}(\mathbf{x},\rho)$ formed from rows $\mathcal{R}(\mathbf{x})\backslash 1$ and columns $\mathcal{Q}(\mathbf{x})$. Then we have the equations
\begin{align}
(\varphi(n,\mathbf{x}))_{n\in\mathcal{R}(\mathbf{x})\backslash 1}={\rm diag}(\widehat{\mathbf{r}}(\mathbf{x}))^{-1}\widehat{\mathbf{A}}(\mathbf{x},\rho){\rm diag}(\widehat{\mathbf{q}}(\mathbf{x}))\widehat{\boldsymbol{\pi}}(\mathbf{x}),
\label{ide1}
\end{align}
for which the left hand side is identified. Since $\mathbf{A}(\mathbf{x},\rho)$ has full rank, $\widehat{\mathbf{A}}(\mathbf{x},\rho)$ has rank $|\mathcal{Q}(\mathbf{x})|$. This is because, when $\rho=1$, we have $\mathcal{R}(\mathbf{x})=\mathcal{Q}(\mathbf{x})$ and $\widehat{\mathbf{A}}(\mathbf{x},\rho)=\mathbf{I}_{|\mathcal{Q}(\mathbf{x})|}$.  When $\rho\in(0,1)$, $\mathcal{R}=\{1,...,\overline{n}(\mathbf{x})\}$, hence $\widehat{\mathbf{A}}(\mathbf{x},\rho)$ comprises rows $\{2,...,\overline{n}(\mathbf{x})\}$ and a subset of columns $\{2,...,\overline{n}(\mathbf{x})\}$ of $\mathbf{A}(\mathbf{x},\rho)$. The sub-matrix of $\mathbf{A}(\mathbf{x},\rho)$ with rows $\{2,...,\overline{n}(\mathbf{x})\}$ and columns $\{2,...,\overline{n}(\mathbf{x})\}$ has determinant $\rho^{T(\overline{n}(\mathbf{x})-1)}>0$, which implies that $\widehat{\mathbf{A}}(\mathbf{x},\rho)$ has rank $|\mathcal{Q}(\mathbf{x})|$. Moreover, ${\rm diag}(\widehat{\mathbf{q}}(\mathbf{x}))$ is invertible by construction, hence \eqref{ide1} identifies $\widehat{\boldsymbol{\pi}}(\mathbf{x})$ for all $\mathbf{x}$ in the support of $\mathbf{x}_g$. 

By assumption there are at least three different group sizes in the support of $n_{g_0}$, so $|\cup_{\mathbf{x}}\mathcal{Q}(\mathbf{x})|\geq 3$. This implies that $\pi(n_1),\pi(n_2),\pi(n_3)$ are identified for at least three different group sizes. We conclude by applying the well known result established by \cite{lee07}, \cite{davezies09} and \cite{bramoulle09} that $(\gamma,\delta,\beta)$ are identified when there are at least three different group sizes and $\gamma\beta+\delta\neq 0$. 

\end{proof}

\begin{proof}of Proposition \ref{the2}\\

\noindent Fix $\mathbf{x}_{g_2}=\mathbf{x}$ where $\mathbf{x}$ is some point in its support. Provided that there is variation in the elements of $\mathbf{x}$, the conditional moment
\begin{align}
\mathbb{E}[\overline{y}_{i}|n_{g_1}=n_1,n_{g_2}=n_2,\mathbf{x}_{g_2}=\mathbf{x}]&= \overline{x}_{i}\varphi'(n_1,n_2,\mathbf{x})\label{redEE1}
\end{align}
identifies $\varphi'(n_1,n_2,\mathbf{x})$ for all such $(\mathbf{x},n_1\geq 2,n_2)$ in the support of $(\mathbf{x}_{g_2},n_{g_1},n_{g_2})$ (it is not identified when $n_1=1$ nor when $\mathbf{x}=c\iota_{n_2}$ for some constant $c$ because in these cases $\overline{y}_{i}=\overline{x}_{i}=0$). We obtain
\begin{align}
\varphi'(n_1,n_2,\mathbf{x})=\psi\pi(n_{1})+(1-\psi)\pi(n_{2})\label{ide2}
\end{align}
We now consider the cases $\psi=0$, $\psi=1$ and $\psi\in(0,1)$ separately. If $\psi=0$ then $(\gamma,\delta,\beta)$ are identified because $\gamma\beta+\delta\neq 0$ and $n_{g_2}$ has at least three points in the support of its distribution. If $\psi=1$ then $(\gamma,\delta,\beta)$ are identified because $\gamma\beta+\delta\neq 0$ and $n_{g_1}$ has at least three points in the support of its distribution. The rest of the proof now considers $\psi\in(0,1)$. 

By standard arguments for two-way bi-partite fixed effects models (e.g., \cite{abowd99}), equation \eqref{ide2} identifies $\mu(n_1)=\psi\pi(n_1)+c$ and $\mu(n_2)=(1-\psi)\pi(n_2)-c$ for every pair $(n_1,n_2)$ corresponding to two vertices in a connected component of $\mathcal{G}_{n_{g_1},n_{g_2}}$, where $c$ is an unknown constant. Let the vertices in the connected component respectively correspond to $\mathcal{N}_1$ and $\mathcal{N}_2$, where $\mathcal{N}_1$ is a subset of the support of $n_{g_1}$ and $\mathcal{N}_2$ is a subset of the support of $n_{g_2}$. 

By assumption either $|\mathcal{N}_1|\geq 3$ or $|\mathcal{N}_2|\geq 3$. Without loss of generality, we present the case of $|\mathcal{N}_1|\geq 3$. If the converse holds, the argument is symmetric, switching the roles of the group sizes in the remainder of the proof. Since the component is connected, we have $2\leq n_{1,1}<n_{1,2}<n_{1,3}$ and $2\leq n_{2}$ such that $\{n_{1,1},n_{1,2},n_{1,3}\}\subseteq\mathcal{N}_1$ and $\{n_{2}\}\subseteq\mathcal{N}_2$. Moreover, $\mu(n)$ is identified for $n\in\{n_{1,1},n_{1,2},n_{1,3},n_{2}\}$. For $i\in\{1,2,3\}$, denote
\begin{align}
\Delta_{i}=\mu(n_{1,i})+\mu(n_{2})=\pi(n_{2})+\frac{\psi(\gamma\beta+\delta)(n_{1,i}-n_{2})}{(n_{1,i}-1+\beta)(n_{2}-1+\beta)}\label{ide4}
\end{align}
Then after some algebra, for $k\in\{1,2,3\}$ we have
\begin{align}
\Delta_{i}-\Delta_{k}=\frac{\psi(\gamma\beta+\delta)(n_{1,i}-n_{1,k})}{(n_{1,i}-1+\beta)(n_{1,k}-1+\beta)}\label{ide3}
\end{align}
Now since $\psi\neq 0$ and $\delta+\beta\gamma\neq 0$,
\begin{align}
\Delta\triangleq\frac{\Delta_{1}-\Delta_{2}} {\Delta_{1}-\Delta_{3}}=\frac{(n_{1,1}-n_{1,2})(n_{1,3}-1+\beta)}{(n_{1,1}-n_{1,3})(n_{1,2}-1+\beta)}\label{ide5}
\end{align}
Since $\Delta$ is identified, $\beta$ is identified by \eqref{ide5} if $\Delta(n_{1,1}-n_{1,3})\neq n_{1,1}-n_{1,2}$. Suppose that $\Delta(n_{1,1}-n_{1,3})=n_{1,1}-n_{1,2}$. Then \eqref{ide5} implies $n_{1,2}=n_{1,3}$, but here we have $n_{1,2}<n_{1,3}$, a contradiction. So $\beta$ is identified. Since $\beta$ is identified, $\psi(\gamma\beta+\delta)$ is identified from \eqref{ide3}, hence $\gamma-\delta/(n_2-1)$ is identified using \eqref{ide4}. Now, since the support of $n_{g_2}$ has at least three elements, we also identify $\Delta'=\mu(n_1)+\mu(n_2')$ for some $n_1$ in the support of $n_{g_1}$ and some $n_2'\neq n_2$ in the support of $n_{g_2}$. Note that $n_1$ and $n_2$ may be in a different connected component of $\mathcal{G}_{n_{g_1},n_{g_2}}$. Since $\beta$ and $\psi(\gamma\beta+\delta)$ are identified, $\Delta'$ identifies $\gamma-\delta/(n_2'-1)$. Since, $\gamma-\delta/(n_2-1)$ and $\gamma-\delta/(n_2'-1)$ are identified and $n_2\neq n_2'$, we identify $\gamma$ and $\delta$.

\end{proof}

\noindent \textbf{IV estimation for missing data with known group sizes}\\

\noindent In stacked form, the structural equation is
\begin{align}
\overline{\mathbf{y}}_g=\beta\mathbf{G}_g(n_{g_0})\overline{\mathbf{y}}_g+\delta\mathbf{G}_g(n_{g_0})\overline{\mathbf{x}}_g+\gamma\overline{\mathbf{x}}_g +\overline{\boldsymbol{\epsilon}}_g,\label{rey}
\end{align}
where $\overline{\mathbf{y}}_g=\mathbf{W}_g\mathbf{y}_g$, $\mathbf{W}_g$ is the $n_g\times n_g$ within-specified-group transformation with diagonal elements $1-n_{g}^{-1}$ and off-diagonal elements $-n_{g}^{-1}$ and $\mathbf{G}_g(n_{g_0})$ is $n_g\times n_g$ with diagonal elements equal to zero and off-diagonal elements equal to $(n_{g_0}-1)^{-1}$. We can also rewrite the reduced form (equation \eqref{red3}) in stacked form as
\begin{align}
\overline{\mathbf{y}}_g=(\mathbf{I}_{n_g}-\beta\mathbf{G}_g(n_{g_0}))^{-1}(\gamma\mathbf{I}_{n_g}+\delta\mathbf{G}_g(n_{g_0}))\overline{\mathbf{x}}_g+(\mathbf{I}_{n_g}-\beta\mathbf{G}_g(n_{g_0}))^{-1}\overline{\boldsymbol{\epsilon}}_g,
\end{align}
where $\mathbf{I}_{n}$ is the identity of dimension $n$, hence we have
\begin{align}
\mathbb{E}[\mathbf{G}_g(n_{g_0})\overline{\mathbf{y}}_g|n_g,n_{g_0},\mathbf{x}_g]&=\gamma\mathbf{G}_g(n_{g_0})\overline{\mathbf{x}}_g+(\gamma\beta+\delta)\left(\sum_{j=2}^\infty\beta^{j-2}\mathbf{G}_g(n_{g_0})^j\right)\overline{\mathbf{x}}_g.\label{ret}
\end{align}
Using \eqref{ret}, it is clear that $\mathbf{G}_g(n_{g_0})^2\overline{\mathbf{x}}_g,\mathbf{G}_g(n_{g_0})^3\overline{\mathbf{x}}_g,...$ are instruments for $\mathbf{G}_g(n_{g_0})\overline{\mathbf{y}}_g$, hence we can estimate \eqref{rey} by linear GMM using $\overline{\mathbf{x}}_g,\mathbf{G}_g(n_{g_0})\overline{\mathbf{x}}_g,\mathbf{G}_g(n_{g_0})^2\overline{\mathbf{x}}_g,...$ as instruments for $\overline{\mathbf{x}}_g,\mathbf{G}_g(n_{g_0})\overline{\mathbf{x}}_g,\mathbf{G}_g(n_{g_0})\overline{\mathbf{y}}_g$ \citep{bramoulle09}.

\marginsize{3cm}{3cm}{0.5cm}{0.5cm}
\singlespacing

\begin{table}
\centering \caption{Monte Carlo Results: Missing data}\label{resultsmc1}
{\footnotesize
\begin{tabular}{l| l| c c| c c| c c| c c}
\hline\hline
\multicolumn{10}{c}{\textbf{Contextual effect only} ($\beta=0$ imposed), $N\approx M=8000$ }\\
\hline
Model&&\multicolumn{2}{c|}{Miss-specified (M)}&\multicolumn{2}{c|}{Known (K)}&\multicolumn{2}{c|}{Unknown (U)}&\multicolumn{2}{c}{Unknown-P (U-P)}\\
\hline
$\rho$&&Mean& RMSE&Mean& RMSE&Mean& RMSE&Mean& RMSE\\
\hline
\multirow{2}{*}{1}&$\delta$&0.502	&	0.084	&	0.502	&	0.084	&	0.502	&	0.084	&	0.502	&	0.084

\\
&$\gamma$&1.002	&	0.06	&	1.002	&	0.06	&	1.002	&	0.06	&	1.002	&	0.06

\\

\hline
\multirow{2}{*}{0.9}&$\delta$&0.426	&	0.112	&	0.501	&	0.086	&	0.499	&	0.099	&	0.499	&	0.099

	\\
&$\gamma$&0.971	&	0.07	&	1.002	&	0.061	&	1.001	&	0.07	&	1.001	&	0.07

\\
\hline
\multirow{2}{*}{0.7}&$\delta$&0.328	&	0.198	&	0.504	&	0.097	&	0.502	&	0.153	&	0.5	&	0.151

	\\
&$\gamma$&0.931	&	0.106	&	1.002	&	0.067	&	1.001	&	0.101	&	1	&	0.1

	\\
\hline
\multirow{2}{*}{0.5}&$\delta$&0.264	&	0.266	&	0.498	&	0.112	&	0.518	&	0.257	&	0.504	&	0.238

\\
&$\gamma$&0.904	&	0.144	&	0.998	&	0.077	&	1.007	&	0.158	&	1	&	0.151

	\\
\hline
\multirow{2}{*}{0.3}&$\delta$&0.225	&	0.332	&	0.505	&	0.146	&	0.896	&	1.602	&	0.54	&	0.468

\\
&$\gamma$&0.892	&	0.204	&	1.002	&	0.096	&	1.262	&	2.825	&	1.017	&	0.279

\\
\hline
\multicolumn{10}{c}{\textbf{Contextual effect only} ($\beta=0$ imposed), $N\approx M=1600$ }\\
\hline
\multirow{2}{*}{1}&$\delta$&0.498	&	0.184	&	0.498	&	0.184	&	0.498	&	0.184	&	0.498	&	0.184

	\\
&$\gamma$&0.999	&	0.132	&	0.999	&	0.132	&	0.999	&	0.132	&	0.999	&	0.132

	\\
\hline
\multirow{2}{*}{0.9}&$\delta$&0.425	&	0.201	&	0.498	&	0.188	&	0.5	&	0.22	&	0.499	&	0.219

	\\
&$\gamma$&0.973	&	0.142	&	1.003	&	0.134	&	1.003	&	0.153	&	1.003	&	0.152

	\\
\hline
\multirow{2}{*}{0.7}&$\delta$&0.325	&	0.277	&	0.502	&	0.217	&	0.511	&	0.347	&	0.501	&	0.332

	\\
&$\gamma$&0.926	&	0.189	&	1	&	0.148	&	1.002	&	0.224	&	0.997	&	0.219

\\
\hline
\multirow{2}{*}{0.5}&$\delta$&0.261	&	0.353	&	0.495	&	0.253	&	0.693	&	1.303	&	0.52	&	0.538

\\
&$\gamma$&0.903	&	0.251	&	0.998	&	0.175	&	1.083	&	0.649	&	1.006	&	0.336

\\
\hline
\multicolumn{10}{c}{\textbf{Contextual and endogenous effects, $N\approx M=8000$}}\\
\hline
\multirow{3}{*}{1}&$\beta$&0.095	&	0.772	&	0.095	&	0.772	&	0.095	&	0.772	&	0.095	&	0.772

\\
&$\delta$&0.471	&	0.245	&	0.471	&	0.245	&	0.471	&	0.245	&	0.471	&	0.245

\\
&$\gamma$&1.02	&	0.182	&	1.02	&	0.182	&	1.02	&	0.182	&	1.02	&	0.182

\\
\hline
\multirow{3}{*}{0.9}&$\beta$&0.092	&	0.822	&	0.076	&	0.788	&	0.04	&	0.81	&	0.04	&	0.81

	\\
&$\delta$&0.394	&	0.325	&	0.479	&	0.246	&	0.485	&	0.263	&	0.485	&	0.263

	\\
&$\gamma$&0.989	&	0.174	&	1.018	&	0.187	&	1.008	&	0.193	&	1.008	&	0.193

\\
\hline
\multirow{3}{*}{0.7}&$\beta$&0.133	&	0.899	&	0.069	&	0.795	&	0.048	&	0.875	&	0.052	&	0.876

	\\
&$\delta$&0.269	&	0.48	&	0.483	&	0.253	&	0.481	&	0.322	&	0.478	&	0.322

	\\
&$\gamma$&0.952	&	0.171	&	1.017	&	0.194	&	1.008	&	0.218	&	1.008	&	0.217

	\\
\hline
\multirow{3}{*}{0.5}&$\beta$&0.122	&	0.928	&	0.052	&	0.825	&	0.067	&	0.914	&	0.07	&	0.919

	\\
&$\delta$&0.205	&	0.577	&	0.484	&	0.272	&	0.485	&	0.413	&	0.475	&	0.404

	\\
&$\gamma$&0.923	&	0.188	&	1.01	&	0.202	&	1.015	&	0.269	&	1.012	&	0.262

	\\
	
	\hline
\multirow{3}{*}{0.3}&$\beta$&0.077	&	0.952	&	0.059	&	0.851	&	0.066	&	0.91	&	0.098	&	0.926

	\\
&$\delta$&0.19	&	0.647	&	0.489	&	0.301	&	0.734	&	1.463	&	0.5	&	0.651

	\\
&$\gamma$&0.907	&	0.246	&	1.015	&	0.213	&	1.21	&	2.77	&	1.032	&	0.404

	\\

\hline
\multicolumn{10}{c}{\textbf{Contextual and endogenous effects, $N\approx M=1600$}}\\
\hline
\multirow{3}{*}{1}&$\beta$&0.066	&	0.888	&	0.066	&	0.888	&	0.066	&	0.888	&	0.066	&	0.888

\\
&$\delta$&0.481	&	0.343	&	0.481	&	0.343	&	0.481	&	0.343	&	0.481	&	0.343

\\
&$\gamma$&1.015	&	0.245	&	1.015	&	0.245	&	1.015	&	0.245	&	1.015	&	0.245

	\\
\hline
\multirow{3}{*}{0.9}&$\beta$&0.129	&	0.922	&	0.095	&	0.915	&	0.103	&	0.915	&	0.104	&	0.915

	\\
&$\delta$&0.387	&	0.425	&	0.48	&	0.35	&	0.477	&	0.394	&	0.476	&	0.394

\\
&$\gamma$&1.004	&	0.246	&	1.029	&	0.26	&	1.031	&	0.276	&	1.031	&	0.275

	\\
\hline
\multirow{3}{*}{0.7}&$\beta$&0.035	&	0.95	&	0.034	&	0.906	&	0.006	&	0.943	&	0.009	&	0.943

	\\
&$\delta$&0.32	&	0.527	&	0.495	&	0.377	&	0.515	&	0.499	&	0.504	&	0.487

\\
&$\gamma$&0.939	&	0.255	&	1.009	&	0.263	&	1.008	&	0.335	&	1.004	&	0.326

	\\
\hline
\multirow{3}{*}{0.5}&$\beta$&0.031	&	0.965	&	0.03	&	0.915	&	0.022	&	0.95	&	0.023	&	0.949

	\\
&$\delta$&0.249	&	0.643	&	0.487	&	0.424	&	0.599	&	1.263	&	0.488	&	0.707

\\
&$\gamma$&0.911	&	0.305	&	1.007	&	0.284	&	1.051	&	0.658	&	1	&	0.432

	\\

\hline\hline
\end{tabular}\\
}
\footnotesize{\textbf{Notes:} Based on 1000 replications and reported to 3 decimal places. `Miss-specified' is the miss-specified model which uses the NLS estimator obtained by treating the specified groups as if they were the groups (i.e., based on the moment
$\mathbb{E}[\overline{y}_i|n_g,\mathbf{x}_g]=\overline{x}_i\pi(n_g)$). `Known' is the model in which the group size is known which uses the NLS estimator under Assumption \ref{ass1} (i.e., based on the moment
$\mathbb{E}[\overline{y}_i|n_{g_0},\mathbf{x}_g]=\overline{x}_i\pi(n_{g_0})$). `Unknown' is the model in which the group size is unknown which uses the GMM estimator under Assumption \ref{ass2}. `Unonwn-P' is a modification of U which additionally uses the parametric restriction $n_{g_0}-2\sim {\rm Binomial}(2,\omega)$ to construct the GMM estimator. Models Unknown and Unknown-P are based on the upper bound on the support of $n_{g_0}$ being known to be $\overline{n}=4$. Results are presented for $(\beta,\delta,\gamma)$ only. Their true values are $(0,0.5,1)$. 
} 
\end{table}

\begin{table}[t!]
\centering \caption{Monte Carlo Results: Uncertain groups}\label{resultsmc2}
{\footnotesize
\begin{tabular}{l| l| c c| c c| c c| c c}
\hline\hline
\multicolumn{10}{c}{\textbf{Contextual effect only} ($\beta=0$ imposed), $N=8000$ }\\
\hline
Model&&\multicolumn{2}{c|}{Rooms (R)}&\multicolumn{2}{c|}{Floors (F)}&\multicolumn{2}{c|}{Known (K)}&\multicolumn{2}{c}{Uncertain (U)}\\
\hline
$\psi$&&Mean& RMSE&Mean& RMSE&Mean& RMSE&Mean& RMSE\\
\hline
\multirow{2}{*}{0.8}&$\delta$&0.41	&	0.123	&	0.501	&	0.135	&	0.496	&	0.066	&	0.516	&	0.131

\\
&$\gamma$&0.992	&	0.062	&	1.001	&	0.034	&	0.999	&	0.04	&	1.002	&	0.062

\\

\hline
\multirow{2}{*}{0.6}&$\delta$&0.321	&	0.198	&	0.498	&	0.135	&	0.5	&	0.061	&	0.504	&	0.142

	\\
&$\gamma$&0.983	&	0.064	&	1	&	0.032	&	1	&	0.033	&	1.001	&	0.063

\\
\hline
\multirow{2}{*}{0.4}&$\delta$&0.232	&	0.281	&	0.497	&	0.133	&	0.5	&	0.064	&	0.501	&	0.145

	\\
&$\gamma$&0.973	&	0.067	&	1	&	0.031	&	1	&	0.028	&	1.001	&	0.063

	\\
\hline
\multirow{2}{*}{0.2}&$\delta$&0.144	&	0.366	&	0.499	&	0.129	&	0.497	&	0.08	&	0.503	&	0.145

\\
&$\gamma$&0.964	&	0.071	&	1	&	0.028	&	1	&	0.026	&	1.004	&	0.058

\\

\hline
\multicolumn{10}{c}{\textbf{Contextual effect only} ($\beta=0$ imposed), $N=1600$ }\\
\hline
\multirow{2}{*}{0.8}&$\delta$&0.402	&	0.208	&	0.494	&	0.333	&	0.495	&	0.14	&	0.564	&	0.29

	\\
&$\gamma$&0.983	&	0.132	&	0.997	&	0.077	&	0.997	&	0.088	&	1	&	0.135

	\\
\hline
\multirow{2}{*}{0.6}&$\delta$&0.313	&	0.263	&	0.503	&	0.327	&	0.493	&	0.137	&	0.538	&	0.31

	\\
&$\gamma$&0.975	&	0.134	&	1.001	&	0.075	&	0.996	&	0.073	&	0.999	&	0.135

	\\
\hline
\multirow{2}{*}{0.4}&$\delta$&0.226	&	0.331	&	0.505	&	0.319	&	0.492	&	0.146	&	0.516	&	0.332

	\\
&$\gamma$&0.966	&	0.136	&	1	&	0.071	&	0.997	&	0.063	&	0.999	&	0.134

\\
\hline
\multirow{2}{*}{0.2}&$\delta$&0.134	&	0.409	&	0.507	&	0.308	&	0.499	&	0.174	&	0.503	&	0.357

\\
&$\gamma$&0.955	&	0.139	&	1.001	&	0.065	&	0.999	&	0.058	&	1.006	&	0.13

\\

\hline
\multicolumn{10}{c}{\textbf{Contextual and endogenous effects, $N=8000$}}\\
\hline
\multirow{3}{*}{0.8}&$\beta$&0.041	&	0.814	&	0.127	&	0.683	&	0.054	&	0.46	&	0.055	&	0.799

\\
&$\delta$&0.392	&	0.36	&	0.429	&	0.384	&	0.466	&	0.214	&	0.492	&	0.302

\\
&$\gamma$&0.998	&	0.165	&	1.007	&	0.051	&	1.001	&	0.064	&	1.01	&	0.167

\\
\hline
\multirow{3}{*}{0.6}&$\beta$&0.039	&	0.848	&	0.112	&	0.663	&	0.07	&	0.461	&	0.099	&	0.8

	\\
&$\delta$&0.301	&	0.495	&	0.431	&	0.385	&	0.462	&	0.227	&	0.458	&	0.354

	\\
&$\gamma$&0.989	&	0.142	&	1.004	&	0.049	&	1.004	&	0.049	&	1.013	&	0.143

\\
\hline
\multirow{3}{*}{0.4}&$\beta$&0.042	&	0.873	&	0.115	&	0.679	&	0.072	&	0.491	&	0.106	&	0.769

	\\
&$\delta$&0.205	&	0.64	&	0.427	&	0.394	&	0.458	&	0.256	&	0.446	&	0.379

	\\
&$\gamma$&0.978	&	0.119	&	1.004	&	0.048	&	1.003	&	0.043	&	1.01	&	0.117

	\\
\hline
\multirow{3}{*}{0.2}&$\beta$&0.032	&	0.881	&	0.119	&	0.653	&	0.106	&	0.561	&	0.123	&	0.724

	\\
&$\delta$&0.123	&	0.772	&	0.422	&	0.391	&	0.436	&	0.303	&	0.433	&	0.393

	\\
&$\gamma$&0.969	&	0.099	&	1.004	&	0.044	&	1.004	&	0.042	&	1.011	&	0.093

	\\

\hline
\multicolumn{10}{c}{\textbf{Contextual and endogenous effects, $N=1600$}}\\
\hline

\multirow{3}{*}{0.8}&$\beta$&0.038	&	0.913	&	0.067	&	0.841	&	0.097	&	0.709	&	0.019	&	0.896

\\
&$\delta$&0.393	&	0.449	&	0.465	&	0.61	&	0.444	&	0.343	&	0.559	&	0.428

\\
&$\gamma$&0.995	&	0.228	&	1.001	&	0.098	&	1.002	&	0.123	&	1.005	&	0.239

	\\
\hline
\multirow{3}{*}{0.6}&$\beta$&0.024	&	0.927	&	0.088	&	0.833	&	0.065	&	0.697	&	0.051	&	0.9

	\\
&$\delta$&0.311	&	0.568	&	0.458	&	0.604	&	0.455	&	0.358	&	0.522	&	0.485

\\
&$\gamma$&0.986	&	0.209	&	1.006	&	0.096	&	0.998	&	0.096	&	1.009	&	0.22

	\\
\hline
\multirow{3}{*}{0.4}&$\beta$&0.019	&	0.94	&	0.096	&	0.833	&	0.107	&	0.716	&	0.045	&	0.888

	\\
&$\delta$&0.223	&	0.7	&	0.457	&	0.579	&	0.435	&	0.386	&	0.493	&	0.55

\\
&$\gamma$&0.974	&	0.192	&	1.006	&	0.091	&	1.001	&	0.081	&	1.005	&	0.198

	\\
\hline
\multirow{3}{*}{0.2}&$\beta$&-0.004	&	0.94	&	0.091	&	0.825	&	0.078	&	0.754	&	0.06	&	0.863

	\\
&$\delta$&0.147	&	0.826	&	0.456	&	0.57	&	0.451	&	0.429	&	0.467	&	0.606

\\
&$\gamma$&0.962	&	0.177	&	1.004	&	0.085	&	1.001	&	0.075	&	1.011	&	0.176

	\\

\hline\hline
\end{tabular}\\
}
\footnotesize{\textbf{Notes:} Based on 1000 replications and reported to 3 decimal places. `Rooms' is the model in which the room is treated as if it were the group with NLS estimator based on the moment $\mathbb{E}[\overline{y}_i|n_{g_1},n_{g_2},\mathbf{x}_g]=\overline{x}_i\pi(n_{g_1})$. `Floors' is the model in which the floor is treated as if it were the group with NLS estimator based on the moment $\mathbb{E}[\overline{y}_i|n_{g_1},n_{g_2},\mathbf{x}_g]=\overline{x}_i\pi(n_{g_2})$. `Known' is the model in which is it known whether the group is the room or the floor (i.e., based on the moment
$\mathbb{E}[\overline{y}_i|n_{g_0},\mathbf{x}_g]=\overline{x}_i\pi(n_{g_0})$). `Uncertain' is the model in which it is unknown whether the group is the room or the floor with NLS estimator based on $\mathbb{E}[\overline{y}_i|n_{g_1},n_{g_2},\mathbf{x}_g]=\overline{x}_i[\psi\pi(n_{g_1})+(1-\psi)\pi(n_{g_2})]$. Results are presented for $(\beta,\delta,\gamma)$ only. Their true values are $(0,0.5,1)$. 
} 
\end{table}

\begin{table}[t!]
\centering \caption{Monte Carlo Results: Uncertain groups (FE)}\label{resultsmc3}
\footnotesize{
\begin{tabular}{l| l| c c| c c| c c| c c}
\hline\hline
\multicolumn{10}{c}{\textbf{Contextual effect only} ($\beta=0$ imposed), $N=8000$ }\\
\hline
Model&&\multicolumn{2}{c|}{Rooms (R)}&\multicolumn{2}{c|}{Floors (F)}&\multicolumn{2}{c|}{Known (K)}&\multicolumn{2}{c}{Uncertain (U)}\\
\hline
$\psi$&&Mean& RMSE&Mean& RMSE&Mean& RMSE&Mean& RMSE\\
\hline
\multirow{2}{*}{0.8}&$\delta$&0.41	&	0.123	&	0.847	&	0.375	&	0.496	&	0.066	&	0.516	&	0.131

\\
&$\gamma$&0.992	&	0.062	&	1.3	&	0.303	&	0.999	&	0.04	&	1.002	&	0.062

\\

\hline
\multirow{2}{*}{0.6}&$\delta$&0.321	&	0.198	&	0.756	&	0.292	&	0.5	&	0.061	&	0.504	&	0.142

	\\
&$\gamma$&0.983	&	0.064	&	1.224	&	0.227	&	1	&	0.033	&	1.001	&	0.063

\\
\hline
\multirow{2}{*}{0.4}&$\delta$&0.232	&	0.281	&	0.668	&	0.218	&	0.5	&	0.064	&	0.501	&	0.145

	\\
&$\gamma$&0.973	&	0.067	&	1.149	&	0.153	&	1	&	0.028	&	1.001	&	0.063

	\\
\hline
\multirow{2}{*}{0.2}&$\delta$&0.144	&	0.366	&	0.584	&	0.156	&	0.497	&	0.08	&	0.503	&	0.145

\\
&$\gamma$&0.964	&	0.071	&	1.075	&	0.081	&	1	&	0.026	&	1.004	&	0.058

\\

\hline
\multicolumn{10}{c}{\textbf{Contextual effect only} ($\beta=0$ imposed), $N=1600$ }\\
\hline
\multirow{2}{*}{0.8}&$\delta$&0.402	&	0.208	&	0.841	&	0.488	&	0.495	&	0.14	&	0.564	&	0.29

	\\
&$\gamma$&0.983	&	0.132	&	1.297	&	0.309	&	0.997	&	0.088	&	1	&	0.135

	\\
\hline
\multirow{2}{*}{0.6}&$\delta$&0.313	&	0.263	&	0.762	&	0.43	&	0.493	&	0.137	&	0.538	&	0.31

	\\
&$\gamma$&0.975	&	0.134	&	1.225	&	0.24	&	0.996	&	0.073	&	0.999	&	0.135

	\\
\hline
\multirow{2}{*}{0.4}&$\delta$&0.226	&	0.331	&	0.679	&	0.375	&	0.492	&	0.146	&	0.516	&	0.332

	\\
&$\gamma$&0.966	&	0.136	&	1.151	&	0.169	&	0.997	&	0.063	&	0.999	&	0.134

\\
\hline
\multirow{2}{*}{0.2}&$\delta$&0.134	&	0.409	&	0.596	&	0.328	&	0.499	&	0.174	&	0.503	&	0.357

\\
&$\gamma$&0.955	&	0.139	&	1.076	&	0.102	&	0.999	&	0.058	&	1.006	&	0.13

\\

\hline
\multicolumn{10}{c}{\textbf{Contextual and endogenous effects, $N=8000$}}\\
\hline
\multirow{3}{*}{0.8}&$\beta$&0.041	&	0.814	&	0.224	&	0.558	&	0.054	&	0.46	&	0.055	&	0.799

\\
&$\delta$&0.392	&	0.36	&	0.725	&	0.375	&	0.466	&	0.214	&	0.492	&	0.302

\\
&$\gamma$&0.998	&	0.165	&	1.318	&	0.324	&	1.001	&	0.064	&	1.01	&	0.167

\\
\hline
\multirow{3}{*}{0.6}&$\beta$&0.039	&	0.848	&	0.198	&	0.565	&	0.07	&	0.461	&	0.099	&	0.8

	\\
&$\delta$&0.301	&	0.495	&	0.645	&	0.347	&	0.462	&	0.227	&	0.458	&	0.354

	\\
&$\gamma$&0.989	&	0.142	&	1.237	&	0.244	&	1.004	&	0.049	&	1.013	&	0.143

\\
\hline
\multirow{3}{*}{0.4}&$\beta$&0.042	&	0.873	&	0.17	&	0.602	&	0.072	&	0.491	&	0.106	&	0.769

	\\
&$\delta$&0.205	&	0.64	&	0.571	&	0.347	&	0.458	&	0.256	&	0.446	&	0.379

	\\
&$\gamma$&0.978	&	0.119	&	1.159	&	0.168	&	1.003	&	0.043	&	1.01	&	0.117

	\\
\hline
\multirow{3}{*}{0.2}&$\beta$&0.032	&	0.881	&	0.152	&	0.612	&	0.106	&	0.561	&	0.123	&	0.724

	\\
&$\delta$&0.123	&	0.772	&	0.492	&	0.36	&	0.436	&	0.303	&	0.433	&	0.393

	\\
&$\gamma$&0.969	&	0.099	&	1.082	&	0.094	&	1.004	&	0.042	&	1.011	&	0.093

	\\

\hline
\multicolumn{10}{c}{\textbf{Contextual and endogenous effects, $N=1600$}}\\
\hline

\multirow{3}{*}{0.8}&$\beta$&0.038	&	0.913	&	0.138	&	0.747	&	0.097	&	0.709	&	0.019	&	0.896

\\
&$\delta$&0.393	&	0.449	&	0.785	&	0.62	&	0.444	&	0.343	&	0.559	&	0.428

\\
&$\gamma$&0.995	&	0.228	&	1.31	&	0.331	&	1.002	&	0.123	&	1.005	&	0.239

	\\
\hline
\multirow{3}{*}{0.6}&$\beta$&0.024	&	0.927	&	0.129	&	0.772	&	0.065	&	0.697	&	0.051	&	0.9

	\\
&$\delta$&0.311	&	0.568	&	0.715	&	0.582	&	0.455	&	0.358	&	0.522	&	0.485

\\
&$\gamma$&0.986	&	0.209	&	1.237	&	0.263	&	0.998	&	0.096	&	1.009	&	0.22

	\\
\hline
\multirow{3}{*}{0.4}&$\beta$&0.019	&	0.94	&	0.138	&	0.789	&	0.107	&	0.716	&	0.045	&	0.888

	\\
&$\delta$&0.223	&	0.7	&	0.617	&	0.565	&	0.435	&	0.386	&	0.493	&	0.55

\\
&$\gamma$&0.974	&	0.192	&	1.161	&	0.191	&	1.001	&	0.081	&	1.005	&	0.198

	\\
\hline
\multirow{3}{*}{0.2}&$\beta$&-0.004	&	0.94	&	0.136	&	0.804	&	0.078	&	0.754	&	0.06	&	0.863

	\\
&$\delta$&0.147	&	0.826	&	0.521	&	0.563	&	0.451	&	0.429	&	0.467	&	0.606

\\
&$\gamma$&0.962	&	0.177	&	1.083	&	0.123	&	1.001	&	0.075	&	1.011	&	0.176

	\\

\hline\hline
\end{tabular}\\
}
\footnotesize{\textbf{Notes:} Based on 1000 replications and reported to 3 decimal places. `Rooms' is the model in which the room is treated as if it were the group with NLS estimator based on the moment $\mathbb{E}[\overline{y}_i|n_{g_1},n_{g_2},\mathbf{x}_g]=\overline{x}_i\pi(n_{g_1})$. `Floors' is the model in which the floor is treated as if it were the group with NLS estimator based on the moment $\mathbb{E}[\overline{y}_i|n_{g_1},n_{g_2},\mathbf{x}_g]=\overline{x}_i\pi(n_{g_2})$. `Known' is the model in which is it known whether the group is the room or the floor (i.e., based on the moment
$\mathbb{E}[\overline{y}_i|n_{g_0},\mathbf{x}_g]=\overline{x}_i\pi(n_{g_0})$). `Uncertain' is the model in which it is unknown whether the group is the room or the floor with NLS estimator based on $\mathbb{E}[\overline{y}_i|n_{g_1},n_{g_2},\mathbf{x}_g]=\overline{x}_i[\psi\pi(n_{g_1})+(1-\psi)\pi(n_{g_2})]$. Results are presented for $(\beta,\delta,\gamma)$ only. Their true values are $(0,0.5,1)$. 
} 
\end{table}

\setlength{\tabcolsep}{2pt}

\newpage
\FloatBarrier
\begin{table}[t!]\centering \caption{Summary statistics  } \label{tab:sum}
\begin{tabular}{l ccc@{\hskip 0.4in}ccc}
\hline\hline
& \multicolumn{3}{c}{All firms}& \multicolumn{3}{c}{Firms with $n_{g_0}\leq 50$ }\\
& \multicolumn{3}{c}{$N_0=8448, G_0=755$ }& \multicolumn{3}{c}{$N_0=5599, G_0=732$ }\\
\hline
&Mean&Median&SD&Mean&Median&SD\\
\hline
\addlinespace
\underline{Individual covariates:}	\\	
Quit-to-exit (0/1)		&       0.036&       0&       0.187   &       0.035&       0&       0.185\\

Female (0/1)		 &       0.450&       0&       0.498  &       0.454&       0&       0.498\\

Grad. degree (0/1)       &       0.376&       0&       0.484&      0.322&       0&       0.467\\
				
Age (years)		 	   &      36.493&      34&      10.145&      36.713&      34&      10.812\\
						
Experience (years)	    &       7.799&       6&       6.927&     7.712&       5&       7.183\\

Own fee-weighted caseload (`000 yuan) 	   &      20.130&       3.050&      39.325&       20.758&       3.798&      38.478\\

\addlinespace
\underline{Group F:}	\\	
Peer fee-weighted caseload   &      20.130&      17.888&      16.101 &      20.758&      16.812&      19.210\\		
		
Prop. high-ability 	    &       0.250&     0.226&       0.172&       0.255&       0.231&       0.204\\

Prop. female 	     &       0.450&       0.444&       0.190 &       0.454&       0.462&       0.230\\

Mean age   &      36.493&      36.493&       5.063&      36.713&      35.800&       6.073\\

Mean experience   &       7.799&       7.532&       3.374&     7.712&       7.000&       4.004\\

Prop. grad. degree   &       0.376&       0.402&       0.210&         0.322&       0.324&       0.229\\

\addlinespace
\underline{Group F-C ($age\leq 35$):}	\\
Peer fee-weighted caseload   &      19.532&      14.562&      18.519  &      20.221&      14.634&      21.684\\		

Prop. high-ability &       0.244&       0.217&       0.193&           0.251&       0.200&       0.228\\

Prop. female 	     &       0.537&       0.538&       0.218&       0.531&       0.500&       0.260\\

Peer experience    &      29.736&      29.667&       1.573&      29.665&      29.600&       1.839\\

Peer experience  &       3.998&       4.000&       1.185&       3.884&       3.750&       1.395\\

Prop. grad. degree  &       0.374&       0.385&       0.228&       0.321&       0.333&       0.251\\

\addlinespace
\underline{Group F-C ($age>35$):}	\\
Peer fee-weighted caseload    &      20.514&      18.430&      18.558&      20.957&      15.902&      22.556\\		

Prop. high-ability peers &       0.253&       0.243&       0.220&         0.253&       0.200&       0.268\\

Prop. female 	     &       0.338&       0.303&       0.228&       0.352&       0.333&       0.283\\

Peer age  &      45.190&      43.108&       5.804&     46.301&      44.818&       6.985\\

Peer experience  &      12.763&      12.500&       4.526&   13.002&      12.429&       5.49\\

Prop. grad. degree  &       0.388&       0.438&       0.260&         0.330&       0.286&       0.301\\
							
\hline \addlinespace
\multicolumn{7}{l}{\footnotesize 
\begin{minipage}{0.95\textwidth}
\scriptsize{\textbf{Notes:} Quit-to-exit is the dependent variable of interest, which takes a value of 1 if a lawyer quits the job and exits the local legal practice in 2016, 0 otherwise. Fee-weighted caseload measures lawyers' ability in civil litigation in the prior three years (2013-2015), weighted by the litigation size. We consider two peer groups, F and F-C. Group F means all associate lawyers in the same law firm form the relevant group. Group F-C means lawyers in the same law firm of the same cohort --young or experienced-- form the relevant group, where young (experienced) cohort is defined as below (above) 35. Group variables are calculated using a ``leave-own-out" method, where high-ability is defined as being in the top quartile of fee-weighted caseload ($> 24.44$). }
 \end{minipage}
    }
\end{tabular}\\
\end{table}
\FloatBarrier

\begin{table}[t!]\centering \caption{Peer effects in lawyers' quit-to-exit} \label{tab:app}
\scriptsize{
\begin{tabular}{l c c c c c c c c c c}\hline\hline
\multicolumn{5}{l}{Dep. var: Quit-to-exit (0/1)}\\
&(1) & (2) &(3) &(4) & (5)& (6)& (7)\\
&All &All &All &All &$age\leq35$& $age>35$& All\\
\hline
\underline{Group F:} \\
\addlinespace
$\quad$ Peer fee-weighted caseload   &      0.0001         &                     &                     &                     &                     &                     \\
                    &   (0.00029)         &                     &                     &                     &                     &                     \\
                                        
$\quad$ Prop. high-ability		 &                     &                     &      0.0544\sym{**} &                     &                     &                     \\
                    &                     &                     &   (0.02201)         &                     &                     &                     \\        
                         
$\quad$ Prop. female &     -0.0136         &                     &     -0.0115         &                     &                     &                     \\
                    &   (0.02893)         &                     &   (0.02840)         &                     &                     &                     \\
                    
$\quad$  Peer age    &      0.0042\sym{**} &                     &      0.0046\sym{**} &                     &                     &                     \\
                    &   (0.00180)         &                     &   (0.00181)         &                     &                     &                     \\

$\quad$  Peer experience    &     -0.0052\sym{*}  &                     &     -0.0056\sym{*}  &                     &                     &                     \\
                    &   (0.00295)         &                     &   (0.00295)         &                     &                     &                     \\
                    
$\quad$ Prop. grad. degree          &     -0.0457\sym{*}  &                     &     -0.0423         &                     &                     &                     \\
                    &   (0.02715)         &                     &   (0.02704)         &                     &                     &                     \\
\addlinespace
\underline{Group F-C:} \\
\addlinespace                    
$\quad$ Peer fee-weighted caseload                   &                     &      0.0003         &                     &                     &                     &                     \\
                    &                     &   (0.00028)         &                     &                     &                     &                     \\
                    
$\quad$ Prop. high-ability   &                     &                     &                     &      0.0451\sym{***}&      0.0767\sym{***}&      0.0211         \\
                    &                     &                     &                     &   (0.01747)         &   (0.02878)         &   (0.02276)         \\        

$\quad$ Prop. female &                     &     -0.0163         &                     &     -0.0142         &     -0.0250         &     -0.0090         \\
                    &                     &   (0.02166)         &                     &   (0.02178)         &   (0.03430)         &   (0.02576)         \\
                    
$\quad$ Peer age     &                     &      0.0008         &                     &      0.0010         &      0.0008         &      0.0007         \\
                    &                     &   (0.00198)         &                     &   (0.00196)         &   (0.00524)         &   (0.00213)         \\

$\quad$ Peer experience        &                     &     -0.0001         &                     &     -0.0004         &     -0.0016         &     -0.0002         \\
                    &                     &   (0.00179)         &                     &   (0.00178)         &   (0.00582)         &   (0.00197)         \\
                                        
$\quad$ Prop. grad. degree    &                     &     -0.0365\sym{*}  &                     &     -0.0347\sym{*}  &     -0.0376         &     -0.0279         \\
                    &                     &   (0.01895)         &                     &   (0.01875)         &   (0.02865)         &   (0.02301)         \\

\addlinespace
\underline{Group U:} \\
\addlinespace                    
                    
$\quad$ Prop. high-ability   &                     &                     &                     &    &    &  &   0.0581\sym{**}          \\
                    &                     &                     &                     &         &         &  &(0.0285)     \\        

$\quad$ Prop. female &                     &             &                     &            &         &    &      -0.0130         \\
                    &                     &           &                     &         &      & &   (0.0285)     \\
                    
$\quad$ Peer age     &                     &           &                     &           &           &   &  0.0021           \\
                    &                     &            &                     &         &        & &  (0.0024)      \\

$\quad$ Peer experience        &                     &            &                     &          &         &  &  -0.0015      \\
                    &                     &      &                     &      & &   & (0.0025)      \\
                                        
$\quad$ Prop. grad. degree    &                     &      &                     &     &        &&   -0.0432\sym{*}      \\
                    &                     &       &                     &        &          &&    (0.0242)   \\
\quad$\psi$&&          &          &            &            &            & 0.5618           \\
&&          &          &            &            &            & (0.6090)           \\
                                        
\addlinespace

Fee-weighted caseload       &     -0.0003\sym{***}&     -0.0003\sym{***}&     -0.0003\sym{***}&     -0.0003\sym{***}&     -0.0003\sym{***}&     -0.0002\sym{***}&-0.0003\sym{***}\\
                    &   (0.00005)         &   (0.00006)         &   (0.00005)         &   (0.00005)         &   (0.00007)         &   (0.00008)& (0.0001)         \\

Female 	    &      0.0146\sym{***}&      0.0128\sym{**} &      0.0148\sym{***}&      0.0129\sym{**} &      0.0159\sym{*}  &      0.0086    &0.0134\sym{**}      \\
                    &   (0.00555)         &   (0.00602)         &   (0.00552)         &   (0.00602)         &   (0.00860)         &   (0.00700)& (0.0061)         \\

Age	  &     -0.0035\sym{**} &     -0.0044\sym{**} &     -0.0035\sym{**} &     -0.0043\sym{**} &      0.0086         &     -0.0001&-0.0043\sym{**}             \\
                    &   (0.00143)         &   (0.00215)         &   (0.00143)         &   (0.00215)         &   (0.01618)         &   (0.00300) &(0.0021)        \\         										
Experience     &      0.0016         &      0.0020         &      0.0016         &      0.0020         &      0.0063         &      0.0008  &0.0019          \\
                    &   (0.00143)         &   (0.00141)         &   (0.00142)         &   (0.00140)         &   (0.00419)         &   (0.00150)  &(0.0014)       \\

Grad. degree    &     -0.0050         &     -0.0066         &     -0.0048         &     -0.0065         &     -0.0133\sym{*}  &      0.0042&-0.0065            \\
                    &   (0.00552)         &   (0.00558)         &   (0.00551)         &   (0.00556)         &   (0.00689)         &   (0.00893)& (0.0055)         \\

\hline
$N_0$                   &        8448         &        8137         &        8448         &        8137         &        4622         &        3515 & 8137        \\
$G_0$ 	& 	755&        1359& 	755&        1359   & 691&669&1359  \\
FE&F&F-C&F&F-C&F-C&F-C&F-C\\
\hline \addlinespace
\multicolumn{8}{l}{\footnotesize 
\begin{minipage}{0.95\textwidth}
\scriptsize{ ***: significant at the 0.01 level, **:0.05 level, *: 0.1 level. \\
\textbf{Notes:} Each column represents a regression; quadratic terms in age and experience are included in the regression. All reported regressors are defined as in Table 4 and Section 7. Standard errors in parentheses are clustered by law firm. Peer group F means all lawyers in the same firm are peers. F-C means lawyers in the same firm of the same cohort (young or experienced) are peers. `U' means there is group uncertainty between F-C and F, with probability $\psi$ of being the former.}
 	 \end{minipage}
    }
\end{tabular}
}
\end{table}

\FloatBarrier
\newpage
\begin{table}[t!]\centering \caption{Peer effects in lawyers' quit-to-exit: Sub-sample of firms with $n_{g_0}\leq 50$ } \label{tab:appnl50}
\scriptsize{
\begin{tabular}{l c c c c c c c c c c}\hline\hline
\multicolumn{5}{l}{Dep. var: Quit-to-exit (0/1)}\\
&(1) & (2) &(3) &(4) & (5)& (6)& (7)\\
&All &All &All &All &$age\leq35$& $age>35$& All\\
\hline
\underline{Group F:} \\
\addlinespace
$\quad$ Peer fee-weighted caseload     &      0.0001         &                     &                     &                     &                     &                     \\
                    &   (0.00031)         &                     &                     &                     &                     &                     \\
                    
$\quad$ Prop. high-ability		&                     &                     &      0.0573\sym{**} &                     &                     &                     \\
                    &                     &                     &   (0.02318)         &                     &                     &                     \\           
                    
$\quad$ Prop. female  &     -0.0150         &                     &     -0.0132         &                     &                     &                     \\
                    &   (0.03082)         &                     &   (0.03001)         &                     &                     &                     \\
                    
$\quad$  Peer age    &      0.0046\sym{**} &                     &      0.0050\sym{**} &                     &                     &                     \\
                    &   (0.00193)         &                     &   (0.00193)         &                     &                     &                     \\

$\quad$  Peer experience     &     -0.0059\sym{*}  &                     &     -0.0062\sym{*}  &                     &                     &                     \\
                    &   (0.00330)         &                     &   (0.00331)         &                     &                     &                     \\
                    
$\quad$ Prop. grad. degree        &     -0.0249         &                     &     -0.0217         &                     &                     &                     \\
                    &   (0.03011)         &                     &   (0.03014)         &                     &                     &                     \\
\addlinespace
\underline{Group F-C:} \\
\addlinespace                    
$\quad$ Peer fee-weighted caseload    &                     &      0.0004         &                     &                     &                     &                     \\
                    &                     &   (0.00033)         &                     &                     &                     &                     \\
                    
$\quad$ Prop. high-ability  &                     &                     &                     &      0.0496\sym{***}&      0.0760\sym{**} &      0.0288         \\
                    &                     &                     &                     &   (0.01835)         &   (0.03012)         &   (0.02333)         \\       

$\quad$ Prop. female  &                     &     -0.0165         &                     &     -0.0149         &     -0.0091         &     -0.0374         \\
                    &                     &   (0.02279)         &                     &   (0.02288)         &   (0.03562)         &   (0.03034)         \\
                    
$\quad$ Peer age        &                     &      0.0007         &                     &      0.0009         &     -0.0023         &      0.0009         \\
                    &                     &   (0.00210)         &                     &   (0.00203)         &   (0.00495)         &   (0.00230)         \\

$\quad$ Peer experience         &                     &      0.0001         &                     &     -0.0001         &      0.0034         &     -0.0006         \\
                    &                     &   (0.00202)         &                     &   (0.00199)         &   (0.00570)         &   (0.00215)         \\
                                        
$\quad$ Prop. grad. degree   &                     &     -0.0201         &                     &     -0.0188         &     -0.0106         &     -0.0181         \\
                    &                     &   (0.02235)         &                     &   (0.02236)         &   (0.03298)         &   (0.02634)         \\
                                        
\addlinespace
\underline{Group U:} \\
\addlinespace                    
                    
$\quad$ Prop. high-ability   &                     &                     &                     &    &    &  &  0.0643\sym{*}             \\
                    &                     &                     &                     &         &         &  &(0.0379)     \\        

$\quad$ Prop. female &                     &             &                     &            &         &    &     -0.0084           \\
                    &                     &           &                     &         &      & &   ( 0.0308)     \\
                    
$\quad$ Peer age     &                     &           &                     &           &           &   & -0.0005               \\
                    &                     &            &                     &         &        & &  (0.0025)      \\

$\quad$ Peer experience        &                     &            &                     &          &         &  &  0.0020         \\
                    &                     &      &                     &      & &   & (0.0033)      \\
                                        
$\quad$ Prop. grad. degree    &                     &      &                     &     &        &&   -0.0505          \\
                    &                     &       &                     &        &          &&    (0.0372)   \\
\quad$\psi$&&          &          &            &            &            & 0.6732               \\
&&          &          &            &            &            & (0.8046)           \\
                                        
\addlinespace

Fee-weighted caseload     &     -0.0003\sym{***}&     -0.0002\sym{**} &     -0.0002\sym{***}&     -0.0002\sym{***}&     -0.0002\sym{**} &     -0.0001  &-0.0002\sym{***}           \\
                    &   (0.00008)         &   (0.00010)         &   (0.00007)         &   (0.00007)         &   (0.00011)         &   (0.00012)         &(0.0001)\\

Female 	     &      0.0137\sym{**} &      0.0120\sym{*}  &      0.0138\sym{**} &      0.0120\sym{*}  &      0.0219\sym{**} &     -0.0075 & 0.0112\sym{*}          \\
                    &   (0.00691)         &   (0.00726)         &   (0.00683)         &   (0.00725)         &   (0.00976)         &   (0.01162)         &(0.0066)\\

Age		 &     -0.0024         &     -0.0032         &     -0.0024         &     -0.0031         &      0.0159         &      0.0043  &  -0.0023       \\
                    &   (0.00177)         &   (0.00219)         &   (0.00177)         &   (0.00219)         &   (0.01952)         &   (0.00314)& (0.0021)        \\     										
Experience	 &      0.0005         &      0.0010         &      0.0005         &      0.0010         &      0.0033         &     -0.0016  &0.0010           \\
                    &   (0.00143)         &   (0.00148)         &   (0.00142)         &   (0.00146)         &   (0.00560)         &   (0.00175)   &(0.0014)      \\  
Grad. degree 	&      0.0027         &      0.0011         &      0.0029         &      0.0011         &     -0.0004         &      0.0096    &-0.0013       \\
                    &   (0.00821)         &   (0.00926)         &   (0.00822)         &   (0.00928)         &   (0.01109)         &   (0.01343)  &( 0.0087)       \\

\hline
$N_0$                    &        5599         &        5288         &        5599         &        5288         &        3083         &        2205&5288         \\
$G_0$ 	& 732 &   1313& 732 &   1313 &  668 & 645    &1313\\
FE&F&F-C&F&F-C&F-C&F-C&F-C\\
\hline \addlinespace
\multicolumn{8}{l}{\footnotesize 
\begin{minipage}{0.95\textwidth}
\scriptsize{ ***: significant at the 0.01 level, **:0.05 level, *: 0.1 level. \\
\textbf{Notes:} Each column represents a regression; quadratic terms in age and experience are included in the regression. All reported regressors are defined as in Table 4 and Section 7. Standard errors in parentheses are clustered by law firm. Peer group F means all lawyers in the same firm are peers. F-C means lawyers in the same firm of the same cohort (young or experienced) are peers. `U' means there is group uncertainty between F-C and F, with probability $\psi$ of being the former.}
 	 \end{minipage}
    }
\end{tabular}
}
\end{table}

\FloatBarrier
\newpage
\begin{table}[t!]\centering \caption{Peer effects in lawyers' quit-to-exit: Robustness } \label{tab:approb}
\scriptsize{
\begin{tabular}{l @{\hskip 0.3in}c @{\hskip 0.2in}c @{\hskip 0.2in}c @{\hskip 0.2in}c @{\hskip 0.2in}c@{\hskip 0.2in} c c c c c}\hline\hline
\multicolumn{5}{l}{Dep. var: Quit-to-exit (0/1)}\\
&(1) & (2) &(3) &(4) & (5)& (6)\\
 \cmidrule(r){2-3} \cmidrule(r){4-5}  \cmidrule(r){6-7}
& \multicolumn{2}{c}{$<55$ subjects only} & \multicolumn{2}{c}{All law firms in} &\multicolumn{2}{c}{Use number of}  \\
&&& \multicolumn{2}{c}{ Shanghai}  &  \multicolumn{2}{c}{ high-ability peers} &\\
\hline

\underline{Group F:}\\
\addlinespace
$\quad$Prop. high-ability   &      0.0573\sym{**} &                          &      0.0733\sym{***}&                   &                     \\
                    &   (0.02474)         &                          &   (0.02238)         &                 &                     \\             
\addlinespace
$\quad$Number of high-ability      &                     & &                     &                     &      0.0202\sym{***}&                     \\
                   &                     &     &                     &                     &   (0.00503)         &                     \\
                    
\underline{Group F-C:}\\
\addlinespace
$\quad$Prop. high-ability       &                     &      0.0509\sym{***}&                          &      0.0421\sym{**}                 &                     \\
                    &                     &   (0.01946)         &                              &   (0.01656)            &                     \\          
\addlinespace
$\quad$Number of high-ability       &                     &   &                     &                     &                     &      0.0200\sym{***}\\
                 &                     &       &                     &                     &                     &   (0.00545)         \\                          
\addlinespace            			
\hline
$N_0$         &        7909         &        7605     &       13613         &       13205                   &        8448         &        8137         \\
$G_0$	 &   	639&     1243 &      1061     &     1919       & 	755&        1359 &       \\
\hline
FE&F&F-C&F&F-C &F&F-C \\
\hline \addlinespace
\multicolumn{7}{l}{\footnotesize 
\begin{minipage}{\textwidth}
\scriptsize{ ***: significant at the 0.01 level, **:0.05 level, *: 0.1 level. \\
\textbf{Notes:}  Each column represents a regression. Odd-numbered models are estimated based on model (3) of Table 5 (group F), and even-numbered models are estimated based on model (4) of Table 5 (group F-C). Models (1)-(2) focus on subjects below age 55. Models (3)-(4) add an interaction term of prop. high-ability and an indicator identifying low-to-medium ability lawyers (caseload below median). Models (5)-(6) add an interaction term of prop. high-ability and an indicator for female. }
 	 \end{minipage}
    }
\end{tabular}
}
\end{table}
\FloatBarrier
\pagebreak

\FloatBarrier
\newpage
\begin{table}[t!]\centering \caption{Peer effects in lawyers' quit-to-exit: Heterogeneity } \label{tab:apphet}
\scriptsize{
\begin{tabular}{l @{\hskip 0.2in}c @{\hskip 0.2in}c @{\hskip 0.2in}c @{\hskip 0.2in}c @{\hskip 0.2in}c@{\hskip 0.2in} c c c c c}\hline\hline
\multicolumn{5}{l}{Dep. var: Quit-to-exit (0/1)}\\
&(1) & (2) &(3) &(4) & (5)& (6)\\
 \cmidrule(r){2-3} \cmidrule(r){4-5}  \cmidrule(r){6-7}
 & \multicolumn{2}{c}{Interacts w/ low-} &\multicolumn{2}{c}{Interacts w/ female}& \multicolumn{2}{c}{Interacts w/ small}  \\
 & \multicolumn{2}{c}{ to med-ability subjects} && & \multicolumn{2}{c}{firms} \\
\hline

\underline{Group F:}\\
\addlinespace
$\quad$Prop. high-ability    &      0.0128         &                     &      0.0693\sym{***}&                     &      0.0642         &                     \\
                    &   (0.02440)         &                     &   (0.02656)         &                     &   (0.04751)         &                     \\                
\addlinespace
$\quad$Prop. high-ability  $\times$               &     0.0698\sym{***}&                     &                     &                     \\
$\qquad $ (Low- to med-ability)                            &   (0.01693)       &                     &                     &                     \\
\addlinespace
$\quad$Prop. high-ability $\times$                 &                     &                     &     -0.0310         &                     \\
 $\qquad $                Female            &                     &                     &   (0.02863)         &                     \\
\addlinespace
$\quad$Prop. high-ability $\times$ Female     &                     &                     &                     &                     &     -0.0109         &                     \\
  $\qquad $ Small firms    &                     &                     &                     &                     &   (0.05194)         &                     \\
 \addlinespace                                                     
\underline{Group F-C:}\\
\addlinespace
$\quad$Prop. high-ability       &                     &      0.0064         &                     &      0.0539\sym{***}&                     &      0.0431\sym{*}  \\
                    &                     &   (0.02008)         &                     &   (0.02071)         &                     &   (0.02235)         \\        
\addlinespace
$\quad$Prop. high-ability  $\times$  &                           &       0.0740\sym{***}&                     &                     \\
$\qquad $ (Low- to med-ability)                 &                     &   (0.01824)          &                     &                     \\
\addlinespace
$\quad$Prop. high-ability  $\times$                 &                     &                     &                     &     -0.0118         \\
$\qquad $ Female           &                     &                     &                     &   (0.02707)         \\                                 
\addlinespace
$\quad$Prop. high-ability $\times$ &                     &                     &                     &                     &                     &      0.0110         \\
  $\qquad $ Small firms                  &                     &                     &                     &                     &                     &   (0.03312)         \\
\addlinespace            			
\hline
$N_0$                     &        7909         &        7605         &        8448         &        8137         &        8448         &        8137         \\
$G_0$ &   	639&     1243 & 	755&        1359 &   	755&        1359    \\
\hline
FE&F&F-C&F&F-C &F&F-C \\
\hline \addlinespace
\multicolumn{7}{l}{\footnotesize 
\begin{minipage}{\textwidth}
\scriptsize{ ***: significant at the 0.01 level, **:0.05 level, *: 0.1 level. \\
\textbf{Notes:}  Each column represents a regression. Odd-numbered models are estimated based on model (3) of Table 5, and even-numbered models are estimated based on model (4) of Table 5. Models (1)-(2) add an interaction term of prop. high-ability and an indicator identifying low-to-medium ability lawyers (caseload below median). Models (3)-(4) add an interaction term of prop. high-ability and an indicator for female. Models (5)-(6) add an interaction term of prop. high-ability and an indicator for small firms (firm size below median). }
 	 \end{minipage}
    }
\end{tabular}
}
\end{table}
\FloatBarrier
\pagebreak

\end{document}